\documentclass[a4paper,11pt]{article}
\pdfoutput=1 
\usepackage{jcappub} 
\usepackage[T1]{fontenc}
 \usepackage{amssymb,amsfonts,slashed,amsthm,amsmath,graphicx}
\usepackage[all]{xy}
\usepackage{dcolumn}
\usepackage{bm}
\usepackage{mathrsfs}
\usepackage{ucs}
\usepackage[utf8x]{inputenc}
\usepackage{hyperref}
\usepackage{hepunits}
\usepackage{slashed}
\usepackage{floatrow}
\usepackage{subfigure}
\usepackage{indentfirst}

\newcommand{\be}{\begin{equation}}
\newcommand{\ee}{\end{equation}}
\newcommand{\bea}{\begin{eqnarray}} 
\newcommand{\eea}{\end{eqnarray}}
\newcommand{\ft}[2]{{\textstyle\frac{#1}{#2}}}

\def\pd{\ensuremath{\partial}}

\def\rme{{\mathrm e}}
\def\rmi{{\mathrm i}}

\newsavebox{\uuunit}
\sbox{\uuunit}
    {\setlength{\unitlength}{0.825em}
     \begin{picture}(0.6,0.7)
        \thinlines
        \put(0,0){\line(1,0){0.5}}
        \put(0.15,0){\line(0,1){0.7}}
        \put(0.35,0){\line(0,1){0.8}}
       \multiput(0.3,0.8)(-0.04,-0.02){12}{\rule{0.5pt}{0.5pt}}
     \end {picture}}

\newcommand{\U}{\mathop{\rm {}U}}

    \def\cL{{\cal L}}
  \def\cN{{\cal N}} \def\cO{{\cal O}} \def\cP{{\cal P}}

  \def\cT{{\cal T}}

\def\atH0{|_{H_{0}}}
\def\AtH0{\bigg|_{H_{0}}}

\def\Nm{\tilde{N}}
\def\Lm{\tilde{L}}
\def\tm{\tilde{t}}

\def\rrm{\tilde{r}}

\def\nmu{\zeta}

\csname @addtoreset\endcsname{equation}{section}

\title{\boldmath {\LARGE{Supermassive Cosmic String Compactifications \vspace{-2.5 cm}}}}
\author[a,b,c]{{\large Jose J. Blanco-Pillado},}
\affiliation[a]{Department of Theoretical Physics, University of the Basque Country UPV/EHU,\\
48080 Bilbao, Spain}
\affiliation[b]{IKERBASQUE, Basque Foundation for Science,  48011, Bilbao, Spain}
\affiliation[c]{Institute of Cosmology, Department of Physics and Astronomy, 
Tufts University, Medford, MA  02155, USA}
\author[a]{{\large Borja Reina,}}
\author[a]{{\large Kepa Sousa}}
\author[a]{{\large and Jon Urrestilla}}

\emailAdd{josejuan.blanco@ehu.es}
\emailAdd{borja.reina@ehu.es}
\emailAdd{kepa.sousa@ehu.es}
\emailAdd{jon.urrestilla@ehu.es}

\abstract{The space-time dimensions transverse to a static straight cosmic string with a sufficiently large tension  (\emph{supermassive cosmic strings}) are compact and typically have a singularity at a finite distance form the core.  In this paper, we discuss how the presence of multiple supermassive cosmic strings in the $4d$ Abelian-Higgs model can induce the spontaneous compactification of the transverse space and explicitly construct solutions where the gravitational background becomes regular everywhere. We discuss the embedding of this model in $\mathcal{N}=1$ supergravity and show that some of these solutions are half-BPS, in the sense that they leave unbroken half of the supersymmetries of the model.}

\begin{document}
\maketitle
\flushbottom

\section{Introduction}
\label{sec:intro}

The gravitational properties of cosmic strings where first studied by Vilenkin   \cite{Vilenkin:1981zs}, who calculated the gravitational background around an  infinitely thin cosmic string in the weak field approximation. He found that the space-time around a cosmic string was conical, that is, Minkowski space minus a wedge,
\be
ds^2 = - dt^2 + dz^2 + dr^2 + (1 -  \ft{\Delta}{2 \pi} )^2 r^2 d \varphi^2,
\label{conicalST}
\ee
where the corresponding deficit angle is proportional to $\mu$, the tension of the string, namely, $\Delta = \mu M_p^{-2}$. This approach had two important limitations. The first problem has to do with the existence of a conical singularity at the position of the string,  that is $r=0$, which is a result of having modeled the cosmic string by a source of zero thickness. Secondly the weak field approximation is only justified for small values of the string tension, indeed the metric above is ill-defined for values of the tension close to the Planck scale $\mu \sim 2 \pi M_p^2$. However, both problems are resolved when  the metric is calculated by considering the full Einstein-scalar-gauge  field equations for the gravitating Abelian-Higgs model. In this case the infinitely thin cosmic string is replaced by an Abrikosov-Nielsen-Olesen vortex \cite{Nielsen:1973cs} whose finite size core smooths out the conical singularity. If the tension of the string is low enough, it is possible to show that the deficit angle is approximately proportional to the vortex tension $\Delta \approx \mu M_p^{-2}$, and that  the metric far from the string core behaves as (\ref{conicalST}) \cite{Garfinkle:1985hr,LagunaCastillo:1987cs}.\\

These are not the only solutions of the Abelian-Higgs  model coupled to gravity. In fact this theory admits a rich variety of cosmic string solutions depending on the choice of parameters in the lagrangian as well as the boundary conditions imposed on the various fields. For instance, depending on the asymptotic behavior of the metric it is possible to find two different families of solutions for each choice of parameters: the so-called \emph{Higgs branch}, which includes the cosmic string solutions with the behavior found by Vilenkin with an asymptotically conical space-time, and the \emph{Kasner branch}  \cite{Laguna:1989rx,Christensen:1999wb,Meierovich:2001sx}, where the metric has an asymptotic behavior of the Kasner type, as in the case of self gravitating static global $\U(1)$ cosmic strings \cite{Gregory:1988xc}.  One could also consider time dependent solutions where the space-time along the string is expanding. Solutions of this type have been consider in global \cite{Gregory:1996dd} as well as local cosmic strings 
\cite{Cho:1998xy} where the intrinsic geometry of the worldsheet is assumed to be an expanding deSitter space, in other words the space-time is inflating along the longitudinal directions to the string. In this paper we will focus on constructions of static configurations based on Higgs branch solutions only and therefore we will not consider the Kasner branch or inflating strings.\\

The deficit angle of a unit winding cosmic string solution in the Higgs branch can be calculated in terms of the parameters of the theory, i.e. the square of the ratio of  the scalar and gauge field masses in the vacuum, $\beta$, and the symmetry breaking scale measured in Planck units, $\gamma$. Depending on the value of the deficit angle, $\Delta(\gamma,\beta)$, we can find three different types of cosmic string solutions in the Higgs branch \cite{VilenkinShellard}: 
  \begin{itemize}
\item For values of $\gamma$ and $\beta$ such that $\Delta(\gamma,\beta) < 2 \pi$, the asymptotic behavior of the metric is the one found by Vilenkin, this type of strings are  called \emph{subcritical  cosmic strings}.

\item   When  $\Delta(\gamma,\beta) = 2 \pi$ the space-time transverse to the string compactifies on a cylinder $\mathbb{M}_4 \to \mathbb{M}_3 \times S^1$ and therefore the conical picture for the space-time has to be abandoned. These solutions are known as \emph{critical cosmic strings}. 
  
\item  For even larger values of the deficit angle $\Delta(\gamma,\beta) > 2 \pi$, the space-time far from the string core has the geometry of an inverted cone, that is, as we go away from the core the length of a circle centered on the string decreases, and eventually, at finite distance from the core it becomes of zero size. At this point the metric usually has a conical singularity \cite{Ortiz:1990tn}. Thus the space-time compactifies as $\mathbb{M}_4 \to \mathbb{M}_2 \times \cT$, where $\cT$ is a closed $2-$dimensional manifold. These strings are known as \emph{supermassive cosmic strings}. 
  \end{itemize}

A natural question to ask is whether the conical singularity appearing far away from the core of a supermassive cosmic string configuration can be smoothed out in a similar way as was done 
for the $r=0$ region in the solution found by Vilenkin (\ref{conicalST}). Actually the conical singularity is consistent with the presence of a second cosmic string of zero thickness with a tension $\mu_s$ \cite{Gott:1984ef}
\be
\mu_s = 4 \pi M_p^2 - \mu,\qquad \Longleftrightarrow \qquad \Delta_s + \Delta = 4 \pi,
\label{singularityDeficit}
\ee 
 and thus we might wonder if it is possible to regularize the metric replacing the singularity by a  second Abrikosov-Nielsen-Olesen.  From this discussion it can be seen that the study of regular supermassive cosmic string solutions leads us to consider static multi-vortex  \nopagebreak configurations\footnote{It has also been suggested that the singularities might be a result of trying to impose the configuration to be time independent, and that they might disappear relaxing this condition \cite{Cho:1998xy}. We will discussed the relation of our configurations to these other inflating solutions in a future publication \cite{next}.}.\\

The possibility of finding regular spontaneous compactifications in the spacetime of a supermassive string was introduced by Gott \cite{Gott:1984ef} who described
the idea of attaching a compact transverse manifold to the strings by appropriately choosing the total deficit angle induced by them. A particularly simple solution of this
type was obtained by Linet \cite{Linet:1990fk} within  the Einstein-Abelian-Higgs model in the Bogomolnyi-Prasad-Sommerfeld (BPS) limit, $\beta=1$. Using the first order
equations of motion found by Gibbons \emph{et al.} in \cite{Comtet:1987wi},  he was able to construct a $2-$vortex configuration with a regular space-time by matching two critical string solutions, that is, both with the same tension corresponding to a  deficit angle of $2 \pi$. \\

After reviewing the existing results in the literature we construct in this paper a new type of regular configurations by matching pairs of parallel supermassive cosmic strings solutions with arbitrary windings.  In order to patch the pairs of vortex solutions we will require the space-time to be regular everywhere and that all observable quantities are continuous across the boundary between the patches. We shall show that the deficit angles $\Delta_1$ and $\Delta_2$ of the vortices used in the construction must satisfy 
\be
\Delta_1 + \Delta_2 = 4 \pi,  \qquad  \Delta_1 \ge \Delta_2 \ge 0,
 \ee 
meaning that only the pairings \emph{supermassive-subcritical} and  \emph{critical-critical}  are possible. 
We explore the parameter space of these types of solutions and identify the possible arrangements of these multi-vortex solutions always looking for solutions with axial symmetry. In some extreme cases, the identity of the individual vortices disappears in a background of constant magnetic field. One can recognize these solutions as the $4$D analogue of
the well known flux compactification discussed in the literature \cite{RandjbarDaemi:1982hi,Salam:1984cj}. Our results offer  quite a different interpretation of these
solutions as the result of squeezing together vortices on a compact space in the gravitating Abelian-Higgs model. Conversely one can also think of
these new multivortex solutions as deformations of the spherical flux compactifications previously studied.\\

Some of the solutions we present in this paper bear a strong resemblance with constructions in higher dimensional space times, 
like the $6$D compactification described in \cite{Salam:1984cj, Aghababaie:2003wz,Aghababaie:2003ar} or the stringy cosmic strings found in \cite{Greene:1989ya} and 
specially to the supersymmetric configuration studied in \cite{Parameswaran:2005mm}.  However, our solutions are somewhat 
different in the fact that they are perfectly smooth geometries that correspond to natural extensions of the
supermassive string configurations. It would be very interesting to see if one could uplift these solutions
to a higher dimensional case preserving the nice properties of our solutions in a more realistic compactification
model. We will consider this matter in a future publication \cite{next}.\\

The  structure of the paper is the following. In section \ref{abelianHiggsSection} we review the  gravitating Abelian-Higgs model, and after giving the  ansatz for the Abrikosov-Nielsen-Olesen vortices, we describe the general properties of the singular compactification induced by a  supermassive cosmic string. 
In sections \ref{matchingSection} and \ref{GeneralProperties} we  discuss the matching procedure, and we give the main properties of our new regular supermassive compactifications. Section  \ref{sphericalCompactificationSection} is dedicated to review a regular spherical flux compactification of the gravitating Abelian-Higgs model,  not involving a cosmic string configuration.  
We present in section \ref{numericalResults} the numerical solutions for all the regular compactifications described earlier and explain how one can obtain them in different limits of the parameter space of the theory. In section   \ref{sugraSection} we discuss the embedding of these solutions  in $\cN=1$ supergravity and their supersymmetric properties. Finally, we relegate some of the more technical discussions to the Appendices.

\section{The Abelian-Higgs model. Gravitating Nielsen-Olesen vortices }
\label{abelianHiggsSection}

The action of the gravitating Abelian-Higgs model  describes the dynamics of a complex scalar field $\phi$, which is minimally coupled to a $U(1)$ gauge boson $A_\mu$:
\begin{equation}
S = - \int d^4 x \sqrt{\mid g\mid } \left( \frac{1}{2} M_p^{2}
{ R} + D_{\mu} \bar\phi 
D^{\mu}\phi
+
{1\over 4 g^2}{F}_{\mu \nu}{F}^{\mu \nu} + {\lambda^2 \over 2} ( \xi- q \bar \phi \phi )^2  \right)
\label{higgsaction}
\end{equation}
where, ${F}_{\mu \nu}=\pd_\mu A_\nu - \pd_\nu A_\mu$ is the abelian field strength and $D_\mu = \partial _{\mu} - iqA_{\mu}$ is the usual gauge covariant derivative (See Appendix \ref{conventions} for the conventions ). With these definitions the theory becomes invariant under $\U(1)$ gauge transformations $\phi \to e^{\rmi q \alpha}\phi$ and   $A_\mu \to A_\mu + \partial_\mu \alpha$, where the gauge parameter $\alpha$ is defined to have period $2 \pi$, and thus the charge $q$ must be an integer.   The lagrangian also involves the constant $g$, which denotes the gauge coupling, and $\lambda$, which is the self coupling of the scalar field. However it is easy to show that the classical dynamics of the  gravitating Abelian-Higgs model can be characterized completely by the two dimensionless parameters $\beta = \lambda^2 / g^2$ and $\gamma = \xi/(q M_p^2)$.  In cylindrical coordinates $(r,\varphi,z)$ the usual ansatz for a winding $n$ cosmic string is

\begin{eqnarray}
\phi=\sqrt{\ft{\xi}{ q}} f(r)e^{i n\varphi}, \qquad \quad
A_\mu dx^\mu = \ft1q v(r)  \, d\varphi.
\label{NOansatz}
\end{eqnarray}

This ansatz represents a static cylindrically symmetric field configuration invariant under boosts along the axis of symmetry which, without loss of generality,  we have taken along the $z-$direction.  The most general line element consistent with these symmetries is 
\begin{equation}
ds^{2} = N^{2}(r) ( -dt^{2} + dz^{2}) + d{r}^{2} + L^{2}(r)d{\varphi}^2.
\label{lineelement}
\end{equation}

For convenience we will measure all lengths with respect to the scale $l_g\equiv 1/(g\sqrt{\xi q})$, thus we perform the reescalings
 \be
r \to l_g \, r, \qquad \qquad L(r) \to l_g \, L(r).
\label{reescalings}
 \ee

Then, the profile functions for the gauge and scalar fields must satisfy the equations 
\be
\frac{(N^2 Lf')'}{N^2 L} - \frac{(n-v)^2}{L^2} \,f +\beta \left(1-f^2\right) f= 0,\qquad 
\frac{L}{N^2}\left(\frac{N^2 v'}{L}\right)' + 2 f^2 (n-v) = 0,
\label{systemNO1}
\ee
and the  $tt-$  and $\varphi\varphi-$components of the Einstein's equations read respectively 
\begin{eqnarray}
\frac{(LNN')'}{N^2 L} &=&\gamma \left( \frac{v'^2}{2L^2} -  \frac{\beta}{2}
(1-f^2)^2\right),
\nonumber \\
\frac{(N^2 L')'}{N^2 L}
&=& -\gamma \left( \frac{v'^2}{2 L^2} +2 \frac{(n-v)^2 f^2}{L^2} + \frac{\beta}{2}
(1-f^2)^2\right).
\label{systemE2}
\end{eqnarray}

Configurations representing isolated cosmic strings satisfy the boundary conditions \cite{VilenkinShellard}:
\begin{eqnarray}
f(0)&=&0, \qquad  \qquad \qquad v(0)=0,\qquad \label{boundaryConditions1} \\
\lim_{r\rightarrow \infty} f(r)& =& 1, \qquad \qquad \lim_{r\rightarrow \infty} v(r) = n.
\label{boundaryConditions}
\end{eqnarray}

Moreover, in order to ensure that  the metric is regular  at the center of the cosmic string, we  impose that    the line element  approaches that of  Minkowski space-time for $r\to0$, that is  $L(r) \approx r$ and $N(r)\approx1$, thus
 \be
 L(0)=0, \qquad  L'(0) = 1, \qquad N(0) = 1,  \qquad N'(0) = 0. 
\label{initialConditions}
 \ee

\begin{figure}[t]
\vspace{-1cm}
\hspace{-.9cm}
\begin{minipage}[]{0.5\linewidth}{
\centering
	\includegraphics[width=\textwidth]{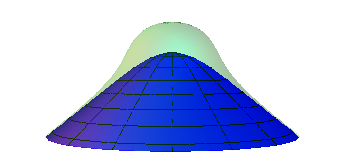}
\vspace{1.0cm}
 }
\end{minipage}
\hspace{.8cm}
\begin{minipage}[]{0.45\linewidth}{
\centering
	\includegraphics[width=\textwidth]{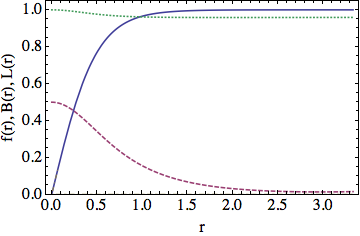}
}
\end{minipage}
\vspace{-0.7cm}
\caption{\footnotesize{   Space-time geometry and profile functions  for a subcritical cosmic string of winding number $n=1$ with  $\beta=11.1$, $\gamma=0.19$ and $\Delta = 0.66 \pi$. LEFT: The lower (blue) surface represents the embedding diagram of the spatial metric on the directions transverse to the string, with the string ($r=0$) placed at the top of the figure. The distance between the lower and the upper (green) surfaces represents the magnetic field density on the transverse space. RIGHT: Scalar condensate  $f(r)$ (solid blue line), magnetic field $B(r)$, reescaled from $B(0)=2.03$ to $1/2$ at $r=0$ to fit nicely in the plot (dashed red), and metric profile function $N(r)$ (dotted green). }}
 \label{subcriticalSM}
\end{figure}

A numerical solution with these boundary conditions can be easily found for the subcritical case, meaning $\Delta < 2 \pi$. We see in Fig. \ref{subcriticalSM}
how the distribution of matter fields smooths out the core of the solution quickly approaching Vilenkin's approximate
solution Eq. (\ref{conicalST}) with $\Delta = 0.66 \pi$.\\

 In the Bogomolnyi limit $\beta=1$ the problem of finding cosmic strings solutions simplifies considerably. It  is  possible to  show that the system of equations (\ref{systemNO1}-\ref{systemE2})  admits a first integral, leading to a new system of first order differential equations called Bogomolnyi equations, which for $n>0$ read:
\be
f' - L^{-1} (n-v) \, f \ = \ 0, \qquad  v' - L (1-f^2) 
\ =
\ 0, \qquad  L' - 1+  \gamma \left( f^2 (n-v) +v \right) =\ 0,
\label{Bog4}
\ee
while the profile function $N(r)$ can be consistently set to a constant, which can be chosen to be $1$ without loss of generality. The energy per unit length $\mu$ of these configurations  saturates the following  bound,  which holds for $\beta\ge1$ \cite{VilenkinShellard}
\be
\mu \, M_p^{-2}\equiv \gamma  \int drd\varphi L \left(f'^2+{(n-v)^2 f^2\over L^2} 
+
{v'^2 \over 2  L^2 }   + {\beta \over 2} (1-f^2 )^2  \right) \ge 2 \pi |n| \gamma.
\label{higgsenergy}
\ee

As we mentioned earlier, the gravitational properties of cosmic strings can be characterized by the deficit angle $\Delta$, which in terms of the fields reads
\be
\Delta = \mu \, M_p^{-2} +2 \pi \int_0^{\infty} dr L (\log N)'^2, \qquad \qquad \Delta =2 \pi(1- \lim_{r\to \infty} L'(r)) ,
\label{deficitTolman0}
\ee
and is related to the asymptotic behavior of the metric profile function $L(r)$  far from the core  $r \to \infty$.  \\

A useful identity can be obtained integrating the difference of the Einstein equations\footnote{This identity reduces to the third of the BPS equations (\ref{Bog4}) in the Bogomolnyi limit $\beta=1$.} (\ref{systemE2})
\be 
\frac{d}{dr}\left[ N^2 L \left( \frac{N'}{N} - \frac{L'}{L} + \gamma\frac{ (n- v) v'}{L^2}\right) \right] = 0.
\label{EinsteinIntegral}
\ee

Evaluating the resulting  expression at $r=0$ and at $r \to \infty$ we can find a relation between the deficit angle, the gravitational potential at infinity $N(\infty)$ and  the magnetic field $B(r)\equiv v'(r)/L(r)$ at the string centre, $B_0$
\be
 \frac{\Delta}{2 \pi} = 1 - \frac{1 - \gamma n \, B_0}{N(\infty)^2}.
\label{BNrelation}
\ee

In the BPS limit the deficit angle takes a very simple form since $N(r)=1$, and from (\ref{Bog4}) it is easy to see that $B_0=1$, therefore:
\be
\Delta= 2 \pi |n| \gamma.
\ee

\subsection{Singular compactification induced by a supermassive string}
\label{singularCompactification}

   In a supermassive cosmic string configuration the space transverse to the  string, $\mathcal{T}$, becomes closed, and the metric assumes the form of a warped compactification, with the gravitational potential $N(r)$ acting as a warp factor
\be
ds^2 = N^2(r) \eta_{\mu \nu} dx^{\mu}dx^{\nu}+ ds^2_\mathcal{T}. 
\label{compactification}
\ee

Here $x^{\mu}$ are now the coordinates in the reduced two-dimensional Minkowski space-time, i.e. $\mu,\nu \in \{t,z\}$, and $\mathcal{T}$, which is parametrized by $r$ and  $\theta$, plays the role of the internal manifold. \\

The condition $\Delta=2 \pi$ defines the boundary between cosmic strings with the usual conical space-time metric (\ref{conicalST}) and those where $\mathcal{T}$ is closed. For a winding  $n$ string this condition  defines a curve on the parameter space $(\beta,\gamma)$ of the Abelian-Higgs model
 \be
\Delta_n(\beta, \gamma)= 2 \pi \qquad \Longrightarrow \qquad \gamma=\gamma_n(\beta).
\label{criticalGamma}
 \ee

In the case  of a  winding one cosmic string this line can be fitted to the following power law $\gamma_1(\beta) = 0.99 \,  \beta^{-0.275}$ \cite{Christensen:1999wb}. As mentioned in the Introduction, supermassive cosmic strings are those which have  a deficit angle larger than $2 \pi$, which implies that the derivative of the profile function $L(r)$ is negative for large values of $r$, and therefore   $L(r)$ must become zero at some finite distance from the core, $r^*$, rendering the transverse space $\mathcal{T}$ closed. In general the transverse space has a  conical singularity at $r=r^*$ with an associated deficit angle $\Delta_s = 4 \pi - \Delta$. If the deficit angle $\Delta$ is close to the critical value,  $\Delta_n(\beta, \gamma)\gtrsim 2 \pi$, the conical singularity occurs far away from the string core and the boundary conditions (\ref{boundaryConditions})  can still be used for $r \to r^*$ \cite{Ortiz:1990tn,Christensen:1999wb}, however for larger values of $\Delta$ they 
 have to be replaced by \cite{Meierovich:2001sx}
 \be
f'(r^*) = 0, \qquad \qquad v(r^*) = n.
\label{alternativeBC}
 \ee

\begin{figure}[t]
\vspace{-2cm}
\centering
 \includegraphics[width=0.35\textwidth]{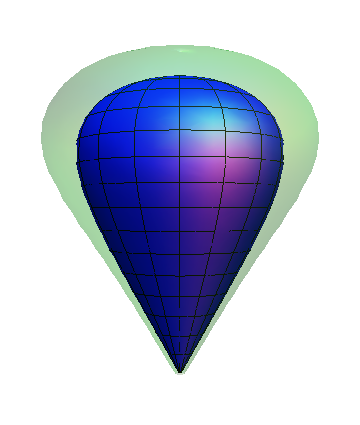}
  \hspace{2.2cm}
 \includegraphics[width=0.45\textwidth]{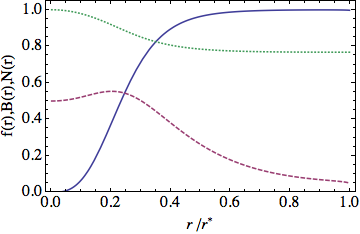}
  \caption{\footnotesize{ Space-time geometry and profile functions for a supermassive cosmic string of winding $n=3$ with  $\beta=11.1$ and $\gamma=0.19$ ($\Delta =2.84 \pi$).     LEFT:  The inner (blue)  surface represents the embedding diagram of the spatial metric on the directions transverse to the strings. The   cosmic string is at the top of the figure ($r=0$), and the conical singularity at the bottom ($r^* = 3.9$). The distance between the inner and the outer (green) surfaces represents the magnetic field density on the transverse space. RIGHT: Scalar condensate  $f(r)$ (solid blue line), magnetic field $B(r)$  (dashed red), and metric profile function $N(r)$ (dotted green). The radial coordinate is measured in units of $r^*$. As in the previous plot the magnetic field is rescaled from $B(0)=2.2$ to $1/2$ at $r=0$.}}
\label{singularSM}
\end{figure}

Indeed, as it is shown  in Fig. \ref{singularSM}, in supermassive configurations with very large deficit angle the magnetic field is non-zero at the conical singularity at  $r^*$. For convenience in the plots representing the profile functions we measure the radial coordinate in units of $r^*$, which also gives the maximum extent for each solution. Notice that, unlike flat space cosmic strings, the maximum of the magnetic field is not at the center of the string. This is related to the fact that, for supermassive cosmic strings with $\beta>1$, the gravitational potential $N(r)$  decreases away from the center of the string, and the magnetic field $B(r)$ is pulled towards  areas of small $N(r)$, as can be shown  from the second equation in (\ref{systemNO1})
\be
B'(r) \sim -(N'/N) \, B(r).
\ee

This effect competes with the   mutual repulsion between the scalar condensate and the magnetic field \cite{Achucarro:1999it} (the so called \emph{Meissner effect} in superconductivity), resulting  in the concentration of the magnetic flux at the boundary layer between the core, where $N(r)$ is maximised, and the asymptotic regime where the scalar field condenses $f(r)\to 1$. Since in the Bogomolnyi limit $N(r)=1$ everywhere, the warping and this effect disappear altogether as we approach $\beta=1$.\\

Having found the solutions for subcritical and supercritical gravitating strings we now proceed to look for 
more complicated configurations that combine both these types of solutions. We will see in the next section 
that this will allow us to find perfectly smooth solutions for the compact space.

\section{Regular supermassive configurations}

\subsection{Matching cosmic string solutions}
\label{matchingSection}

In this section we will use similar techniques as the ones described in \cite{Gott:1984ef,Hiscock:1985uc} to remove the singularity from the supermassive cosmic string configuration by matching  two regular string solutions. We start by dividing the space-time in two patches which we shall call `North' (N) and `South' (S).  The field configuration on each patch is determined by two sets of  fields  which we will denote by $\{ \phi^{(N)}, A_\mu^{(N)}\}$, and $\{\phi^{(S)}, A_{\tilde \mu}^{(S)}\}$, and the  space-time geometry is   characterized by the corresponding line element of the form (\ref{lineelement})
\bea
ds^2_N&=&N^2(r)( - dt^2+dz^2)+dr^2+L^2(r)d\varphi^2 \\ 
d s^2_S&=&\tilde N^2(\tilde r) (- d \tilde t^2+ d \tilde z^2)+d \tilde r^2+\tilde L^2(\tilde r)d \tilde \varphi^2.
\label{elementNS}
\eea

The coordinates are defined in the  ranges $t \in (-\infty,\infty)$, $\varphi\in[0,2 \pi)$ and $z \in (-\infty,\infty)$ (and similarly for the coordinates in the southern patch). The radial coordinates take values in the intervals  $r \in[0,r_N]$ and $\tilde r \in[0,r_S]$,  and therefore the boundary of the northern and southern patches  are cylinders centered on the strings with radii $r_N$ and $r_S$ respectively. With these definitions the distance between the two strings in the final configuration is $r^* = r_N+r_S$. This quantity is related to the size of the transverse space since, after the matching, the center of one string is the farthest point over $\mathcal{T}$ to the position of the other string.\\  

In order to construct the cosmic string solutions the equations of motion, (\ref{systemNO1}) and  (\ref{systemE2}) have to be solved in each of the two patches,   setting the  winding numbers $n_1$ and $n_2$  for the scalar fields in each patch, and imposing the same boundary conditions at  the cores, $r=0$ and $\tilde r=0$, as for an isolated cosmic string (\ref{boundaryConditions1}). Then, the boundaries are  identified, requiring that  all measurable quantities must be continuous and have continuous derivatives across them. In particular this implies that there is no distribution of energy or charge on the boundaries.\\

The space-time is matched in such a way that in the final configuration the two cosmic strings are parallel to each other. 
This can be done  following the procedure described in the Appendix \ref{matching}, which ensures that the resulting space-time preserves the same symmetries as each individual patch, i.e. the staticity and cylindrical symmetry  around  \emph{both axes}, the one at  $r=0$ and the one at $\tilde r=0$. 
Choosing the coordinates conveniently it is possible to show  that the metric profile functions must be continuous and have continuous derivatives
  \bea
 N(r_N) =& \tilde N(r_S)\,,\qquad \qquad&  L(r_N) = \tilde L(r_S), \nonumber \\
 N'(r_N) =& -\tilde N'(r_S)\,,\qquad \qquad&  L'(r_N) = -\tilde L'(r_S).
\label{metricMatching1}
\eea

Note that for the metric profile function $\tilde N(\tilde r)$ in the South patch we have to relax the conditions at the point $\tilde r=0$, that is, we can only impose $\tilde N'(0)=0$. This normalization at the axis $N(0)=1$  is done rescaling  the coordinate $z$ in the northern patch, but because of the  requirement $N(r_N)=\tilde N(r_S)$  we no longer have the same freedom in the southern patch.\\

Physically the matching  condition on $L(r)$  means that the size of a circle centered on any of the two strings must be the same if we measure it just after or just before the boundary surface. The relative minus sign relating the derivatives between patches is just an artifact of the chosen coordinates, since the radial vectors  in each patch $\hat r_N= \partial_r$ and $\hat r_S=\partial_{\tilde r}$ point in opposite directions at the boundary. \\

The gauge sector is matched requiring that  gauge dependent quantities from both patches are  related by a gauge transformation $A^{(N)} - A^{(S)} =d \alpha^{NS}$,  and $\phi^{(N)} =\phi^{(S)} \rme^{\rmi q \alpha^{NS}}$, while gauge invariant quantities must be continuous at the boundary. The continuity of $\phi$ and the cylindrical symmetry require that the gauge transformation between the two patches is of the form $ \alpha^{NS} = m \theta/q + \alpha_0$, with $m\in \mathbb{Z}$.  In order to be consistent with an the ansatz of the form  (\ref{NOansatz}) in each patch we must also require that $m=n_1+n_2$, and thus  the matching conditions for the scalar and gauge profile functions read
\bea
f(r_N) =& \tilde f( r_S), \qquad   \qquad   &v(r_N)+ \tilde v(r_S) = n_1 +n_2,\label{gaugeMatching1}
 \\
f'(r_N) =& -\tilde f'(r_S), \qquad   \qquad   &v'(r_N)- \tilde v'(r_S)=0.
\label{gaugeMatching2}
\eea

To derive the matching condition for the gauge profile function it is necessary to take into account that the coordinates $\varphi$ and $\tilde \varphi$ run in opposite directions on the boundary layer. Furthermore, it is clear from our construction that the boundary between the two patches does not have any physical significance therefore
its location is arbitrary. For convenience the position of the boundaries $r_N$ and $r_S$ will be defined in the following way 
\be
v(r_N)=n_1, \qquad \qquad \tilde v(r_S)=n_2,
\label{boundary1}
\ee
which is consistent with  (\ref{gaugeMatching1}), and is analogous to  the usual boundary condition for the gauge profile function of an isolated cosmic string (\ref{boundaryConditions}). We shall denote the resulting field configuration by the winding numbers of the matched  cosmic strings $(n_1,n_2)$, and without loss of generality it will be assumed that  $n_1\ge n_2$.

\subsection{General properties of regular supermassive configurations}
\label{GeneralProperties}

 The matching   condition  relating the derivatives of  $L(r)$ (\ref{metricMatching1}) can be rewritten in terms of the deficit angles of the two strings. Indeed, defining the $\Delta_1$ and $\Delta_2$ as the deficit angle in the northern and southern patches respectively we have that
\be
\Delta_1 \equiv 2 \pi (1- L'(r_N)), \qquad \Delta_2 \equiv 2 \pi (1- \tilde L'(r_S)) \quad  \Longrightarrow \quad \Delta_1 + \Delta_2 =4 \pi,
\label{totalDeficit}
\ee
which implies, as we anticipated in the Introduction, that the total deficit angle of the regular supermassive configuration must be $4 \pi$, and that only the matchings \emph{subcritical-supermassive} or  \emph{critical-critical}  are possible. Note that  critical-critical  configurations are necessarily of the form $(n,n)$, since the two matched strings  have the same deficit angle. From the bound on the string tension (\ref{higgsenergy})
and the expression for the deficit angle (\ref{deficitTolman0}) it is possible to derive  an upper limit to the value of $\gamma$ for these configurations
\be
 \Delta = 4 \pi    \ge ( \mu_{1} + \mu_2 ) \, M_p^{-2}\ge 2 \pi  \gamma \,  (n_2 +n_2) \qquad \Longrightarrow \qquad \gamma_{n_1,n_2}  \le \frac{2}{(n_1 + n_2)}, 
\label{gammaBound}
\ee
which is saturated at the BPS limit $\beta=1$. This bound is only applicable when $\beta\ge1$, but we shall argue in later sections that the regular supermassive configurations we are considering can only exist in this regime.\\
   
Moreover, since the  transverse space $\mathcal{T}$  which results after the matching  is compact, the magnetic flux on it, $\Phi_m$, has to be quantized. Indeed, it is easy to see that the matching condition  for the gauge potential translates to  the usual Dirac quantization condition 
\be
\Phi_m = \int_{\mathcal{T}} drd\varphi F_{r \varphi} = \int_{r=r_N} d \varphi \, A_\varphi^{(N)} + \int_{r=r_S} d \tilde \varphi \, A_{\tilde \varphi}^{(S)}   = \frac{2 \pi m}{q}, \quad \text{with} \quad m \equiv n_1 + n_2.
\ee

In the following we will focus on solutions with non-zero $m$. The magnetic field flux quantization will also 
help us to uncover the nature of our solutions. In our previous presentation we have described our configurations as multivortex solutions in a compact space. This characterization may lead 
us to think that the total winding number should be equal to zero, in other words that the type of constructions we are considering would have to involve a vortex and an anti-vortex.   This would certainly be the case in a global $\U(1)$ model, where one would have to have a vortex-anti-vortex configuration
to satisfy the boundary conditions at the intersection layer.  However, this does not have to be true in our model, and in fact
it is never the case in the solutions presented here. To understand why this is so we notice that one can identify the nature of the vortices in the solutions by their 
contribution to the magnetic field over the compact manifold. Having a non-vanishing quantized magnetic field on
this compact space clearly demonstrates the prevalence of one type of vortices over their anti-vortex
counterparts in all our solutions. \\

Once the total flux is identified in our solutions, we still need to specify 
how much of it goes into the North and South regions of the configuration.
In the following we will present several solutions of this type with different
distributions of flux concentrated in each of the poles. For some cases, the interpretation of 
isolated vortices with quantized fluxes would be very clear while in others this clear-cut
distinction would not be so obvious due to the interactions between the vortices. An extreme example
will be described in the following section where we study an spherical compactification with a
uniform magnetic field over the internal space. In this case the vorticity of the scalar field has disappeared altogether and the solution resembles the so-called flux compactification solutions. We will discuss its connection to more general multi vortex solutions in subsequent sections.

\subsection{Spontaneous compactification to $\mathbb{M}_2\times S^2$}
\label{sphericalCompactificationSection}

The results of \cite{Meierovich:2001sx} show that, when the alternative boundary conditions (\ref{alternativeBC}) are taken into account, supermassive cosmic string configurations (even the singular ones) can only exist below a critical value for the gravitational coupling $\gamma \le \gamma^{cr}_{n}(\beta)$, which depends both on the winding number $n$ and the parameter $\beta$.  Indeed, for large values of $\gamma$ the gravitational interaction becomes so strong that it induces the restoration of the  gauge symmetry  everywhere, i.e. $f(r)=0$, and thus, because of the absence of symmetry breaking, cosmic strings cannot form. Moreover, in this regime any configuration with a non-zero magnetic flux over the compact directions  is also a solution of the Einstein-Maxwell system. In this section we will review a simple  analytic solution of the Eintein-Maxwell model which corresponds to  a spherical compactification of the form   $\mathbb{M}_4\to\mathbb{M}_2\times S^2$ \cite{Gott:1984ef,Hiscock:1985uc,Meierovich:2001sx,Som:1994qe}.   \\

 When the Higgs field is set to zero $\phi=0$  the Abelian-Higgs model coupled to gravity  reduces to the Eintein-Maxwell system with constant scalar potential  term $V =\ft12 \lambda^2 \xi^2$ acting as an effective cosmological constant. Requiring the field configuration to be static and cylindrically symmetric about the $z-$axis it is possible to show that the energy per unit length satisfies a BPS type of bound 
\be
\mu \, M_p^{-2}={\gamma \over 2} \int_{\mathcal{T}} drd\varphi L
 \left(L^{-1} \, v'  \pm  \beta^{1/2} \right)^2  \pm   \gamma \beta^{1/2}\int_{\mathcal{T}} drd\varphi \,  v' \ge \pm  2 \pi \, \gamma \beta^{1/2} \, m,
\label{tension}
\ee
with $m$ being the topological number related to the total magnetic flux passing through the transverse space $\mathcal{T}$.  
For field configurations saturating the bound, i.e. if the squared term  vanishes,  the gravitational potential $N(r)$ can be consistently set to a constant, which can be chosen to be one, and thus the compactification is not warped. In that case the constant of motion (\ref{EinsteinIntegral})  translates into a first order differential equation for $L(r)$ \cite{Hartmann:2012ad}.  Using the  ansatz (\ref{NOansatz}) with $f(r)=0$, the resulting differential equations for the gauge  and metric profile functions read 
\be
v' \mp \beta^{1/2}\;L =0, \qquad  \qquad L' = 1 \mp  \gamma \beta^{1/2} \; v,
\label{sphericalBPS} 
\ee
which have to be solved together with the following boundary conditions which  ensure the regularity of space-time metric
\be
L(0)=0, \qquad v(0)=0,  \qquad L'(r^*)=-1, \qquad v(r^*) =m,
\ee
where $r^*$ is the point where the metric profile function vanishes $L(r^*) = 0$. The corresponding solutions are given by:
\be
\phi(r)=0, \qquad  v(r)=   \frac{m}{2}(1  - \cos( r /r_0)), \qquad  L(r) = r_0\sin( r /r_0),  \qquad   N(r)=1,
\label{sphericalCompactification}
\ee
where, after undoing the reescalings (\ref{reescalings}), $r_0=l_g/{\sqrt{\gamma \beta}}$. These solutions can only be found provided the parameters satisfy $\pm m  \gamma \beta^{1/2} = 2$, and therefore  we must choose the signs so that $\pm m = |m|$. Finally, one can also see that the magnetic field is uniform over the transverse space, $B=m/(2\,  r_0^2) $. It is interesting to note that the solutions we present here solve the first order equations of motion even though we have not specified the BPS limit, in other words, there exist $\mathbb{M}_2 \times S_2$ compactications for $\beta \neq 1$. This is due to the fact that one
can construct a purely Einsten-Maxwell supergravity theory (without any scalar field present)  that would give rise to exactly the same equations of motion that we described earlier. We will discuss the relation of our solutions to this supergravity theory in section \ref{sugraSection}, and will be considered in further detail in a subsequent publication \cite{next}.\\

Solutions like these ones were first introduced in a $6d$ context in \cite{Salam:1984cj}. Similarly to what happens in the higher dimensional case, the radius of the sphere is stabilized by
the competition of several ingredients that act as a potential for this radius. In our case, these ingredients are,   the curvature
of the sphere, the constant magnetic field on the sphere (given by a monopole like configuration for the gauge field), and the cosmological
constant (induced by the potential). While the curvature induces  a reduction of the radius, due to flux quantization shrinking the size of the sphere leads to an increase of the magnetic field density, and thus of the magnetic energy.  Taking into account these different contributions one can find compactifications of the form, $AdS_2 \times S_2$ (similar
to the Bertotti-Robinson solutions \cite{Bertotti:1959pf,Robinson:1959ev}), $dS_2 \times S_2$ (similar to Nariai solutions \cite{Nariai}) and our fined tune case, where the spacetime is
described by $\mathbb{M}_2 \times S_2$. This arguments suggest the existence of a lower dimensional
landscape of flux vacua analogous to the $6d$ case recently discussed in \cite{BlancoPillado:2009di}. \\

We shall show in the next section how these spherical solutions can be seen as the limiting case of  more general  configurations with several vortices in a warped $\mathbb{M}_2 \times {\cal T}$ spacetime, with a metric  which is regular everywhere.   \\

\section{Numerical Results}
\label{numericalResults}

In order to solve the coupled system of non-linear differential equations (\ref{systemNO1}), (\ref{systemE2}) we have used a combination of  relaxation and  shooting methods. The relaxation technique is specially relevant to  characterize  how the properties of the solutions depend on  the  parameters of the model. Following the works \cite{Christensen:1999wb,Verbin:2001ye}, we have first discretized the radial coordinate, and then we have solved  the resulting set of non-linear algebraic equations using a multidimensional Newton-Raphson method. The trial solutions needed for the relaxation procedure have to be very close to the final solution for the algorithm to converge, and they have been  constructed using a shooting method.

\subsection{BPS limit: \emph{critical-critical} matching.} 
\label{LinetCompactification}
This particular example of a spontaneous compactification in the Abelian-Higgs model was first found by  Linet in  \cite{Linet:1990fk}. In this work Linet gave  an explicit solution for  critical cosmic strings  at the BPS limit $\beta=1$, and then constructed an $(n,n)$ configuration  by matching two  of such solutions. Moreover, these results also show that the inter-string distance $r^*$  can be set arbitrarily, i.e., it is a zeromode which also has a  one to one correspondence with the size of the scalar condensate at the matching point $f(r_N)=\tilde f(r_S)$. Note that in order to construct this type of configurations the gravitational coupling must saturate the bound (\ref{gammaBound}), giving
\be
 \Delta_{n,n}(1,\gamma)=\Delta_n + \Delta_n = 4\pi \qquad  \Longrightarrow \qquad   \gamma_{n,n}=\frac{1}{ |n|}.
\label{gammaConstraint}
\ee

It is interesting to notice that in this particular case the deficit angle of the final configuration is exactly the sum of the deficit angles of the isolated critical strings before the matching. As we shall see in the next subsection, due to the interactions between the strings  this is not the case in general, however the presence of a zero-mode associated with the inter-string distance indicates that  these cosmic string solutions are non-interacting. This result is closely related  to the string behavior   in flat space, where it is possible to construct static multi-vortex solutions in the BPS limit \cite{VilenkinShellard}, and the vortex positions are also  associated to zero-modes.  \\

\begin{figure}[t]
\vspace{-1cm}
\centering
 \includegraphics[width=0.13\textwidth]{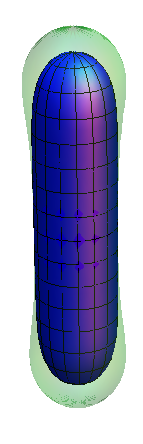}
\hspace{3cm}
  \includegraphics[width=0.48\textwidth]{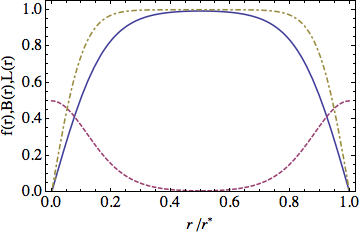}
  \caption{\footnotesize{
  Space-time geometry and profile functions  for a configuration $(1,1)$ with  $\beta=\gamma=1$, and  an inter-string distance is   $r^*=9.4$. LEFT:  Embedding diagram of the transverse space metric (inner blue  surface) and magnetic field density over it (outer green surface). The two strings are located at the top and bottom ends of the figure. RIGHT: Profile functions: scalar field profile function   $f(r)$ (solid blue line), the magnetic field  (dashed red) and $L(r)$ (dot-dashed yellow). The radial coordinate is measured in units of $r^*$, and the magnetic field is rescaled from $B(0)=1$ to $1/2$  at $r=0$. } }   \label{critical6}
\end{figure}

Fig. \ref{critical6} shows the profile functions for a $(1,1)$ configuration with an inter-string distance $r^*=9.2$ at $\beta=1$, and the corresponding embedding diagram  of the space-time metric over $\mathcal{T}$. The transverse space  has a cigar type of geometry with the magnetic field well localized on its ends. Since these solutions occur in the BPS limit the gravitational potential is constant $N(r)=1$ and thus there is no warping in the compactification. \\

We have confirmed  the  presence of the zero-mode numerically; note however,  that we found  a minimum inter-string distance below which no solution can be found. Its value corresponds to the case where $f(r_N)=\tilde f(r_S)=0$, and is given by  $r^*=\pi/\sqrt{ \gamma  }$, half the length of the equator  of the spherical compactification (\ref{sphericalCompactification}) with the same flux, $\Phi_m = 4 \pi |n| /q$, at  $\beta=1$. Actually, in this limit the profile functions become equal to those of  solution (\ref{sphericalCompactification}): the transverse space becomes $S^2$ and the magnetic field turns uniform. We have also used perturbation theory techniques to check the connection between these two configurations (see Appendix \ref{analyticSolutions}), and have shown that exciting a massless mode  of the spherical compactification  with total flux $m=2 n$, one can recover the tubular $(n,n)$ configuration  as a perturbative solution. We thus show that those two solutions, the sphere and the $(n,n)$ configurations, are related, and one is indeed the limiting case of the other. \\

 According to the interpretation given in \cite{Linet:1990fk} the two strings would  be a vortex anti-vortex pair, however our analysis shows that the magnetic field inside both strings must have the same orientation. Indeed,   when the  inter-string distance reaches the minimum, we are left with a configuration with a non-zero magnetic flux over the transverse space $\mathcal{T}$, contrary to what  would be expected with   a vortex anti-vortex pair, which is determined  by the topological number $m=2n$.\\

Moreover, in appendix \ref{matching} we  show that,  in the particular case when we match two critical strings, the gravitational sector is consistent with  a final configuration where the strings move at a constant relative speed $\pd_t \, r^* =const$.  We expect these time dependent configurations to be approximate solutions to the full equations of motion similarly to what happens in the moduli space approximation of critical vortices in flat space. We will of course have to take into account the modifications of the scalar and gauge sectors, but those are expected to be small variations provided the inter string distance is large compared to their core radius. In that case  the fields are very close to their vacuum values at the matching point, and thus the matching conditions in the gauge sector are approximately satisfied.   \\

One may wonder whether it is possible to find configurations of type $(n_1,n_2)$ with different winding numbers $n_1\neq n_2$ in the BPS limit, but it can be shown that the only solutions
to the  first order equations are the spherical solution and the cigar-shaped  compactifications where $n_1=n_2$. There are just no
other cylindrically symmetric solutions compatible with the BPS equations of motion.  Indeed, integrating the BPS equations (\ref{Bog4}) for the gauge and metric profile functions  over the whole volume is straight forward to obtain the following relations\footnote{The equation for the metric profile function needs to be multiplied by $L$ first, and is simplified using that $L (1 -f^2) v = v v' = \ft12(v^2)'$. }
\be
\frac{2 \pi (n_1+n_2)}{A} - 1 +<f^2> =0, \qquad \qquad  \frac{2 \pi (n_1+n_2)}{A} -1 +\gamma n_1 <f^2> =0,
\label{BarlowEq}
\ee
where $A$ is the area of the compact space $\cT$, and average size of the scalar condensate $<f^2>$ is defined as
\be
 <f^2> = \frac{2 \pi}{A}\int_0^{r^*} dr \, L f^2 , \qquad \qquad A =2 \pi \int_0^{r^*} dr \, L.
\ee 

The two previous relations are only compatible when either $<f^2>=0$, and thus $f=0$ everywhere, or when $\gamma n_1=1$. The first possibility corresponds to the compactification on the sphere, and the second can only be a configuration of type $(n,n)$, where $f$ can be non-vanishing, since in the BPS limit the bound for $\gamma$  (\ref{gammaBound}) is saturated.\\

Having identified all the regular cylindrically symmetric supermassive configurations at the critical coupling, a natural question to ask is whether we can find more general solutions  relaxing the condition $\beta=1$. In order to search for these new solutions  we have used perturbation theory techniques described in Appendix  \ref{analyticSolutions} as a guide to look for $(n_1,n_2)$ configurations with beta close to, but different from, the critical one.\\

Perturbing a spherical compactification with the same total flux $m=n_1+n_2$ we have found that, while  perturbative solutions of type $(n,n)$ cannot be constructed when $\beta \neq1$,  it is possible to find  configurations  with $n_1 \neq n_2$ in the case of $\beta>1$ that agree perfectly with fully numerical ones that will be discussed in the next section.  \\

Finally, we have also been able to show that it would be impossible to find perturbative $(n_1,n_2)$ solutions  in the $\beta<1$ regime in agreement with our failed numerical search in this part of the parameter space.  All these analytic checks give us further confidence on the validity of our numerical solutions.

\begin{figure}[t]
\vspace{-1cm}
\centering
 \includegraphics[width=0.35\textwidth]{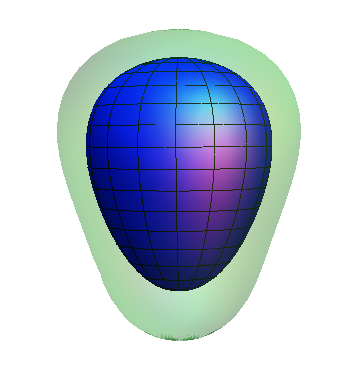}
 \hspace{1.5cm}
 \includegraphics[width=0.48\textwidth]{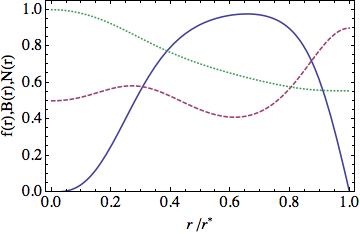}
  \caption{\footnotesize{Space-time geometry and profile functions  for a configuration $(3,1)$ with  $\beta=11.1$ and $\gamma=0.19$.  The conical singularity has been resolved  by a second cosmic string at $r^*=3.3$. LEFT:  Embedding diagram of the transverse space metric (inner blue surface) and magnetic field density over it (outer green surface). The higher and lower winding strings are located at the top and bottom ends of the figure respectively. RIGHT: Profile functions: condensate size $f(r)$ (solid blue line), magnetic field (dashed red), and the gravitational potential $N(r)$ (dotted green). The radial coordinate is measured in units of $r^*$, and the magnetic field is rescaled from $B(0)=1.9$ to $1/2$ at $r=0$.}}
 \label{critical2}
\end{figure}

\subsection{Non-BPS regime} 

\label{numericalProperties}

\subsubsection*{\emph{Supercritical-subcritical} matching.} 

A particular example of the  result  of solving numerically the system of  equations and boundary conditions (\ref{systemNO1} - \ref{initialConditions})    
in each of the patches, together with the matching requirements (\ref{metricMatching1} - \ref{gaugeMatching2}) for a subcritical-supercritical configuration can be seen in Fig. \ref{critical2}. The 
left plot represents the profile functions for a supermassive configuration of type $(n_1,n_2)=(3,1)$  for the choice of parameters $\beta=11.1$ and  
$\gamma=0.19$. This is just the combined configurations
of the $n=1$ subcritical string presented in Fig. \ref{subcriticalSM} with the $n=3$ shown in Fig. \ref{singularSM}. One can see that the full configuration has distorted both parts
of the solution to accomodate our matching conditions. It was reasonable to expect that those solutions presented 
earlier separately could be combined  to a single compact solution since the sum of their deficit angles was comparable to $4\pi$,
the required value for a compact solution. The difference between the sum of  the deficit angles of the isolated strings and $4 \pi$, indicates the existence of a non-trivial interaction between the vortices. Using the fact that  gauge invariant quantities are smooth across the boundary, the profiles of both patches are displayed in a continuous way on a single plot,    with the $n_1-$string  at $r=0$ and the $n_2-$string at a distance $r^*=3.3$ from the first one.\\

 Linearizing the equations of motion around the points  $r=0$ and $\tilde r=0$, it can be shown that the boundary conditions (\ref{boundaryConditions}) and (\ref{initialConditions}) ensure that the scalar and gauge profile functions behave as in the core of isolated flat space cosmic strings with these particular winding numbers, namely, 
\be
f \sim r^{n_1}, \ \ v\sim r^2, \ \ \text{for $r \to 0$,} \qquad \text{and} \qquad
 f \sim \tilde r ^{n_2}, \ \ v\sim \tilde r^2 \ \ \text{for $\tilde r \to 0$}.
\ee

Far from the center of both strings, as we approach the boundary  between the patches, 
the value of the scalar  condensate $|\phi|$ grows and  the $\U(1)$ symmetry is spontaneously broken. The magnetic field concentrates on the cores of the two strings, and although it has a minimum at the point where the scalar field modulus is maximum, it never becomes zero.  Note that the gravitational potential $N(r)$ decreases monotonically towards the lower winding string producing a similar effect to the one observed in singular supermassive strings:
the magnetic field is pulled towards areas of small $N(r)$, so that its density peaks on the $n_2-$string core, and on the $n_1-$string core the  maximum is shifted towards its boundary layer. \\

\begin{figure}[t]
\centering
\includegraphics[width=0.4\textwidth]{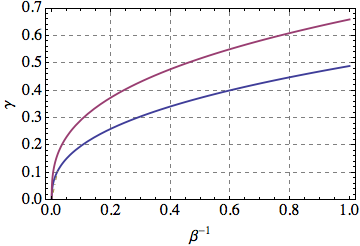}
  \hspace{2.2cm}
 \includegraphics[width=0.4\textwidth]{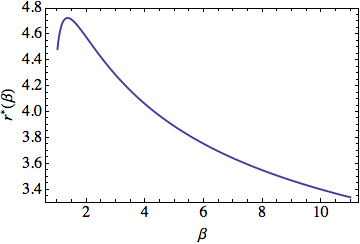}
  \caption{\footnotesize{  LEFT: The lines represent the possible regular supermassive solutions in the $(\gamma, 1/\beta)$ plane for the configurations $(n_1,n_2)=(2,1)$ (bottom) and $(3,1) (top)$.  RIGHT: Inter-string distance $r^*$ (also related to the  size of the transverse space $\mathcal{T}$) as function of beta on a $(3,1)$ configuration.   It decays as $r^*\sim 5.4 \, \beta^{-0.2}$ for  $\beta \to \infty$.}}
 \label{phaseR}
\end{figure}

The geometry of the transverse space $\mathcal{T}$ of the same configuration is represented in the left plot of  Fig. \ref{critical2}, which as expected has the topology of a sphere. The inner (blue) surface is an embedding diagram for the metric  on $\mathcal{T}$, and the distance between the inner and the outer (green) surface represents  the  magnetic field density over it. The $n_1-$string is located at the top of the egg-shaped figure, and the $n_2-$string is at its bottom. Note that, since the core of the higher winding string, where space is approximately flat, is wider than the one of the lower winding one, the curvature around the  $n_1-$string is  lower than around the  $n_2-$string. \\

The qualitative features observed on this figure are generic for  more general choices of the parameters and windings $n_1> n_2$. However, with the boundary  conditions at the center of both  strings together with the matching conditions presented in the previous section,  the profile functions are overdetermined, and in general it will not be possible to find a solution to the system of equations (\ref{systemNO1}-\ref{systemE2})  for any arbitrary value of the constants $\beta$ and $\gamma$. Indeed, our results show that given a value for the parameter $\beta$ it is  possible to construct a $(n_1,n_2)$ type of configuration for only a particular value of  $\gamma$,  such that the total deficit angle is $\Delta_{n_1,n_2}(\beta,\gamma) = 4 \pi$. The corresponding lines of available solutions in the parameter space for string configurations of type $(2,1)$ and $(3,1)$ are displayed on the left plot of  Fig. \ref{phaseR} which have been  fitted to a power law of the form $\gamma_{n_1,n_
 2}(\beta) = c_1 \beta^{c_2}$
 \be
\gamma_{1,2}(\beta) = 0.66 \,  \beta^{-0.35}, \qquad \qquad \gamma_{1,3}(\beta) = 0.49 \, \beta^{-0.39}.
 \ee

\begin{figure}[t]
\centering
\subfigure[ Gravitational potential $N(r)$]{
   \includegraphics[width=0.4\textwidth]{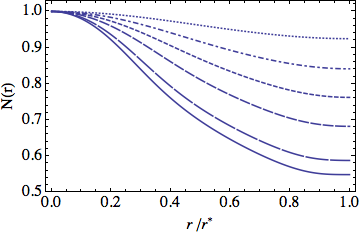}
}
\hspace{2.2cm}
\subfigure[Metric profile function $L(r)$]{
   \includegraphics[width=0.4\textwidth]{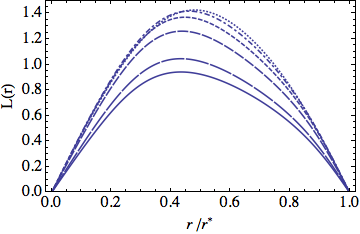}}
\subfigure[Condensate size $f(r)$]{
   \includegraphics[width=0.4\textwidth]{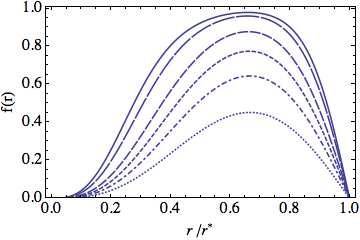}
}
\hspace{2.2cm}
\subfigure[Magnetic field density $B(r)$]{
  \includegraphics[width=0.4\textwidth]{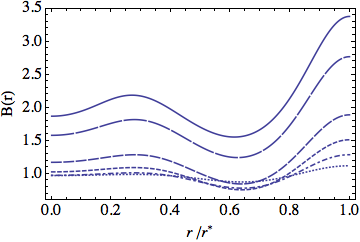} 
}
\caption{\footnotesize{ Profile functions of a $(3,1)$ configuration  for different values of the parameter $\beta$. The different line patterns correspond to the same values of beta  in all the figures. For example, in the upper left plot the  values   of $\beta$  are from bottom to top  $\beta= 11.1, 6.7, 2.7, 1.7, 1.3$, and  $1.1$.}}
 \label{fBetaB}
\end{figure}

All the solutions we have found are in the regime $\beta\ge 1$ and satisfy the bound  for $\gamma$ (\ref{gammaBound}). 
We show in Fig.   \ref{fBetaB} the profiles for the relevant functions for the $(3,1)$ family of solutions characterized by different values of $\beta$ along the curve represented in Fig. \ref{phaseR}.
  In order to show  the dependence  on the inter-string distance, the function  $r^*(\beta)$ associated to these configurations is displayed on the right plot of Fig. \ref{phaseR}. \\

In the limit $\beta\to \infty$, the size of the transverse space contracts since both the inter string distance $r^*$ and the maximum of $L(r)$ (Fig. \ref{fBetaB}), i.e.,  the radius of the largest circle centered on the symmetry axis, decrease for large $\beta$. As the total magnetic flux is  fixed by the topology, this implies that the mean magnetic field on $\mathcal{T}$ must grow (see Fig.  \ref{fBetaB}). At the same time the gradient of the  gravitational potential $N(r)$ (i.e., the warping) becomes more pronounced, and the 
  maximum size of the scalar condensate $|\phi|$ tends to it vacuum expectation value, $\sqrt{\xi/q}$. \\
  
  In the opposite limit,  as one approaches $\beta \rightarrow 1$ the maximum of the scalar profile function decreases to zero, the form of the magnetic field
is uniform over the internal geometry that becomes a perfect sphere, and the gravitational potential $N(r)$ turns into a constant. This limit represents the solution where the two kinds of vortices merge together 
bringing the scalar field to the top of its potential everywhere acting as a pure cosmological constant, that is, the spherical compactification described in section \ref{sphericalCompactificationSection}. As a consequence the value of gamma has to saturate the bound  (\ref{gammaBound}), and thus 
\be
\lim_{\beta\to 1} \gamma_{n_1,n_2} (\beta) = \ft{2}{|n_1+n_2|}.
\label{gammaM}
\ee
\subsubsection*{The critical-critical limit}
 The study of the  extreme case where we have the same amount of flux on both ends of the geometry is numerically very challenging,  in particular  we  cannot reliably say what
is the maximum extent of the geometry.  Our numerical results show that an approximate  $(n,n)$ configuration  seem to exist for the same values of $\gamma$ and $\beta$ as a critical cosmic string with flux $n$ (see eq. (\ref{criticalGamma}))
\be 
\gamma_{n,n}(\beta) \approx \gamma_{n}(\beta).
\ee

This is to be expected in the limit of large inter-string distances in other words when $r^*$ is large compared
to the core size of the vortex. In this case the scalar and gauge field are both approximately in the vacuum at the matching point between the strings, and due to  the symmetry of the configuration all the matching conditions are trivially satisfied. This suggests that we should think of the  critical flux$-n$ string as a limit of the $(n,n)$ configuration when  distance to the second string ($r^*$)  is infinite. However these results can only be trusted up to the finite precision of the numerical calculation, which in the case of gamma gives an uncertainty  of order $10^{-2}$, a much higher value than in all the other numerical solutions described in this paper.\\

\begin{figure}[t]
\centering
 \includegraphics[width=0.32\textwidth]{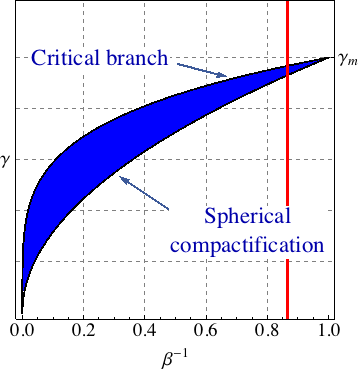}
\hspace{2.cm}
\includegraphics[width=0.45\textwidth]{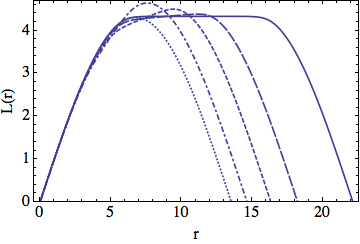}
  \caption{\footnotesize{LEFT: Area covered by the different branches of solutions in the large $m$ limit. The limiting curves correspond to the
spherical compactification and the $(m/2, m/2)$ solution. The vertical line represents the cut through the parameter space that we do to obtain the curves on the right. In the plot we have denoted by $\gamma_m$ the upper bound for the gravitational coupling  (\ref{gammaM}).
RIGHT: The form of the internal geometry, $L(r)$, in the limit of large flux, $m=50$. The dotted line represents
the spherical configuration close to the $(49,1)$ case. We also plot the $(38,12)$, $(29,21)$, $(26,24)$ and $(25,25)$ cases. We see how
the solutions interpolate from the perfectly spherical solution to the tubular configuration of 2 critical strings of half the flux separated
by a straight segment. The distance between the strings increases substantially as one approaches the even split, the  $(25,25)$ case. 
All these curves are computed for $\beta = 1.2$.}}
 \label{Largem}
\end{figure}

On the other hand,  the  perturbative analysis of the equations of motion near the critical coupling  (see Appendix \ref{analyticSolutions}), suggests that  there are no time independent  configurations  of this type when the inter-string distance is small $r^*\to 0$, indicating the presence of short distance interactions between the strings.  It is well known that in flat space it is not possible to find static multi-vortex solutions away from the critical coupling, actually for values of $\beta>1$ vortices repel each other, and for $\beta<1$ they attract. Moreover,  there are indications that vortex solutions of the gravitating Abelian-Higgs model interact in a similar way \cite{Brihaye:2000qr}.    Thus for $\beta<1$ the attraction between vortices would induce a reduction in $r^*$  leading to these solutions to collapse towards  the spherical compactification, and in the case $\beta>1$ the repulsion between vortices would produce a growth of $r^*$, so that $(n, n)$ solutions evolve to two infinitely separated critical strings. An indication that this could indeed be the case is shown in Fig.  \ref{Largem}, where we plot the distance for the static configurations for the different splits for $m=50$ and  $\beta > 1$. We see there that the static configurations become more and more separated as one approaches the critical-critical case in agreement with the idea that there will not be any static finite size configuration in the limiting case of even split of the flux. However, for local vortices these interactions decay exponentially with the inter-string distance, and therefore well separated vortices  only experience an exponentially suppressed force. Thus, we conjecture that, away from the critical coupling, the  configurations $(n, n)$ can only exist as approximate solutions to the static equations of motion for large values of the inter-string distance. An example of such an approximate numerical configuration can be seen in Fig.  \ref{Largem} for the  $(25,25) $ case.

\subsubsection*{The large$-m$ limit}

One can also consider the possibility of having a configuration with a large magnetic field flux over the internal
manifold. In this case we can have many different combinations of the integer flux concentrated
in each end of the internal space. Each of these ways of splitting the total flux would have itself a branch of solutions
for different values of $\beta$ and $\gamma$ similar to what we saw for $m=3,4$ in Fig. \ref{phaseR}. We show in Fig. \ref{Largem}
the distribution of these branches in the $(\gamma, 1/\beta)$ plane for a generic value of $m$. In the case of a large flux, $m$, the
curves describing these solutions cover, in a very dense way, the region between the extreme lines that represent the  $(m-1,1)$ case and the
$(m/2, m/2)$ one.  We show in Fig. \ref{Largem} how the geometry of the solutions interpolate (for a fixed value of $\beta$) between the almost
spherical case, the $(m-1,1)$ solution, to the more tubular elongated form of the $(m/2+1, m/2-1)$ case.
In particular, as we argue in Appendix \ref{largeNAppendix}, in the large$-m$ limit the $(m-1,1)$ configuration 
exists for values of beta and gamma which are precisely  those associated to the spherical compactification with the same total flux  (see section \ref{sphericalCompactificationSection})
\be
\lim_{m\to \infty} \gamma_{m-1,1}(\beta)  = \ft{2}{ |m| } \beta^{-1/2}.
\ee

As one approaches configuration with the even split in the flux,  $(m/2, m/2)$, the solution resembles
the situation with two identical concentrations of vortices whose fields are well separated. The matter fields relax to their vacuum 
form in the central region of the geometry where the metric becomes essentially flat.\footnote{We show in the Appendix \ref{largeNAppendix} the figures with
the form of the matter fields in these configurations and compare them with the analytical approximation in the large $m$ case. The results
show a very good agreement.}\\

In this section, we have shown that there are two different limits in which one can reach a spherical compactification where
the geometry of the internal space is almost perfectly a sphere and the scalar field vanishes.  On the one hand any solution of type $(n_1,n_2)$ approaches the spherical compactification   when  $\beta\to1$, and on the other hand, given a family of total flux $m=n_1+n_2$ and fixing the value of $\beta$, we recover again the same limit  as we approach the solution $(m-1,1)$.     \\

As we mentioned in section \ref{sphericalCompactificationSection} the spherical  compactifications presented there have natural extensions where the geometry of the non-compact dimensions is either deSitter or anti-deSitter space. It would be interesting to think of extensions of these configurations with  supermassive cosmic strings in more dynamical spacetimes such as, a possibly warped, $dS_2 \times {\cal T} $ or $AdS_2 \times {\cal T}$.

\section{Embedding of the solutions in $\cN=1$ supergravity}
\label{sugraSection}

In this section we will discuss the supersymmetric properties of the regular compactifications   discussed in the present article.\\

It is well known that the critically coupled (i.e. with $\beta =1$) Abelian-Higgs model can be embedded in $\cN=1$ supergravity \cite{Dvali:2003zh,Dasgupta:2004dw}.  In its supersymmetric extension the lagrangian describes the dynamics of a graviton multiplet containing the vielbein $\rme_\mu^a$ and its fermionic partner the  gravitino $\Psi_\mu$,  a gauge multiplet which involves the gauge field $A_\mu$ and the gaugino $\lambda$, and   a chiral multiplet consisting of a scalar field $\phi$ and the chiralino $\chi$. The  K\"ahler potential $K(\phi, \bar \phi)$ is chosen so that   the kinetic term is canonical, and the gauged isometry is characterized by the killing vector $k(\phi)$, which corresponds to the usual $\U(1)$ phase shifts:
\be
K(\phi,\bar \phi) = \phi \bar  \phi , \qquad k(\phi) = \rmi q \phi \, \partial_{\phi}.
\ee  

The gauge kinetic function is chosen to be constant $f(\phi) = 1/g$, and since we are interested in supersymmetric cosmic string solutions,  the superpotential is set to zero $W(\phi)=0$. In order to have spontaneous breaking of the gauge symmetry  a Fayet-Iliopoulos term $\xi$ must be included  in the theory, and thus the moment map associated to $k(\phi)$, and the corresponding $D-$term read 
\be
\cP = \xi -q \,  \phi \bar \phi , \qquad D = g^2\left(\xi - q \phi \bar  \phi \right).
\ee

 With these choices the bosonic sector of the action is given by (\ref{higgsaction}), with $\lambda=g$ (i.e. $\beta=1$), and  the same definitions for the covariant derivatives.  In a purely bosonic configuration, the only non-vanishing supersymmetry transformations are the ones of the fermions. The  transformations of the   gravitino $\Psi_\mu$, the chiralino $\chi$ and the gaugino $\lambda$, are given by
\bea
\delta \psi _{\mu L} &=&  (\partial_\mu + \ft14 \omega_\mu^{ab} \gamma_{ab} + \ft{\rmi}{2} A_\mu^B ) \epsilon_L\,, \label{gravitinoSUSYtrans}\\
    \delta \chi_{L} &=& \ft12 \gamma^\mu D_\mu \phi\,  \epsilon_R \,,  \label{chiralSUSYtrans}\\ 
\delta \lambda &=& \ft14 \gamma^{\mu\nu} F_{\phantom{a}\mu\nu} \epsilon + \ft12 i D \gamma_5  \epsilon\,. 
\label{gauginoSUSYtrans} 
\eea

Here $\epsilon$ is the parameter of the supersymmetry transformations, $\omega_\mu^{ab}$ is the spin connection, and $\gamma^\mu$ represent the gamma matrices as usual. The composite field $A_\mu^B$, called the gravitino $U(1)$ connection, is defined in the following way
\be
A_\mu^B = \ft \rmi 2 \left( \bar \phi D_\mu\phi -\phi  D_\mu\bar \phi   \right) - \xi A_\mu.
\ee

 The BPS equations (\ref{Bog4}) ensure that half of the supersymmetries are unbroken provided that the supersymmetry parameter is of the form 
\be
 \epsilon _L(\theta) = \rme^{\mp\ft12\rmi\varphi } \epsilon _{0L}, 
\ee
and $\epsilon _{0}$  satisfies the proyector condition \cite{Dvali:2003zh}
\be  
 \gamma^{12} \epsilon_0 = \mp \rmi \gamma^5 \epsilon_0.
\label{projectorDef}
 \ee
 
The spherical compactification of the Abelian-Higgs (\ref{sphericalCompactification}), and the $(n,n)$ regular supermassive configurations discussed in section \ref{LinetCompactification} are all solutions to the system of BPS equations.  In order for the configuration  to preserve half of the supersymmetries, it must be possible to find a killing spinor that is globally well defined. Since the space-time metric and the field configuration are both regular everywhere, the killing spinor is well defined everywhere\footnote{For the $(n,n)$ regular supermassive string this can easily be checked noting that they are solutions to the BPS equations with the alternative boundary   conditions (\ref{newBoundaryConditions}), introduced in Appendix \ref{analyticSolutions}, and thus there is no need of using the matching.}. Thus we can conclude that \emph{all these configurations leave unbroken half of the supersymmetries} and thus they are also BPS  states  from the point of view of supersymmetry. \\

In $\cN=1$ supergravity, the parameter $\xi$ (i.e., the Fayet-Iliopoulos term) is not arbitrary, it must be an even integer \cite{Distler:2010zg,Seiberg:2010qd} in units of the Plack mass,  $\xi = 2 p M_p^2$, with $p\in \mathbb{Z^+}$. The dimensionless quantity $\gamma$ then reads $\gamma = 2 p/q$.  The spherical compactification discussed here can only exist provided the parameters satisfy the  constraint $|m| \gamma \beta^{1/2} = 2$. Taking into account the quantization of the FI term and that $\beta=1$, the constraint reads 
\be
|m| p= q  \qquad \Longrightarrow \qquad r_0^{-1} =   q \, gM_p, \quad  \text{with} \quad q \in \mathbb{Z}^+,
\label{parameterQuantization}
\ee
 implying that the radius of the spherical compactification also obeys a quantization condition.\\

As we discussed in section  \ref{sphericalCompactificationSection}, the spherical compactification is described by a set of first order differential BPS equations regardless of the value of $\beta$.  This can can be understood recalling that the sphere is a solution to the Einstein-Maxwell system,  which is described by the action (\ref{higgsaction}) with the scalar field set to zero everywhere $\phi=0$. Ignoring the presence of the  chiral multiplet, this action admits an infinite number of supersymmetric embeddings, and in particular, when $g$ is an integer multiple of $\lambda$, it is possible to find one  where the sphere is a BPS configuration in the sense that it preserves half of the supersymmetries. In order to construct such a theory we just have to set $\phi=\chi=0$ and replace the FI term  $\xi$ by $(g  \xi)/\lambda$, (which is also quantized), in the previous construction. 
The fact that the equations of motion admit first integrals is granted by the structure of the supersymmetric action, and the corresponding equations of motion, and is not affected by the  quantization of the FI term which is a topological consistency condition. Thus the spherical compactification can be derived from a set of first order BPS equations for any arbitrary   value of $\lambda \in \mathbb{R}$.  However, when the chiral multiplet containing the scalar field   is taken into account the only allowed supersymmetric embedding where the spherical compactification leaves unbroken half of the supersymmetries is the critically coupled one, $\lambda = g$.\\
  
 In the  spherical and the cigar-shaped compactifications ($n_1=n_2$)  discussed here  the gravitational potential $N(r)$ is constant and therefore there is no warping. The warped compactications discussed in section \ref{numericalResults}, i.e. the $(n_1,n_2)$ supermassive solutions of the Abelian-Higgs model with $n_1\neq n_2$,  can only appear away from the BPS limit $\beta \neq 1$.  Although  the  non-critical Abelian-Higgs model can be embedded in a supersymmetric theory  \cite{Dvali:2003zh}, its cosmic solutions do not satisfy  the BPS equations    (\ref{Bog4}) and therefore they cannot be supersymmetric. Nevertheless, in Appendix  \ref{analyticSolutions} we  argue that for values of $\beta$
 close to one, the size of the scalar condensate is small, of order $\sqrt{\beta- 1}$, and the configuration is very close to the  spherical compactification.  Since the scalar profile function appears in the BPS equations  (\ref{Bog4})  quadratically, this implies that these configurations are still approximate solutions close to the critical coupling, and thus setting $\beta$ sufficiently  close to $1$ the supersymmetry breaking scale can made arbitrarily small.

\section{Discussion} 

In the present work we have analyzed spontaneous compactifications of the four dimensional Abelian-Higgs model down to two-dimensional Minkowski space.  In general the compactification is done on a warped geometry of the form $\mathbb{M}_4 \to \mathbb{M}_2\times \cT$, with the internal manifold having the topology of a two-sphere and a metric which is regular everywhere. These solutions have  a (quantized)  magnetic flux along the compact directions which  might appear trapped inside the core of parallel cosmic strings when the mass of the gauge boson is lower than the mass of the scalar field ($\beta=m_s^2/m_g^2>1$). 
In particular we have focused on configurations having  a cylindrically symmetric internal space, which involve two parallel vortices placed at opposite ends of the compact space. These pairs  can either be formed by one supercritical (deficit angle $\Delta_1> 2\pi$) and one subcritical ($\Delta_2< 2\pi$) string, or two critical ones ($\Delta_1=\Delta_2 = 2\pi$). Below we summarize the possible types of compactifications, denoting  configurations  which involve cosmic strings by the  the corresponding pair of  winding numbers,  $(n_1,n_2)$:  
\begin{itemize}
\item \emph{Spherical compactification, $\cT\cong S^2$}.  This is the four dimensional analog of the Salam-Sezgin spherical compactification \cite{Salam:1984cj,Hiscock:1985uc,Gott:1984ef}.  In this configuration the scalar field is at the top of the potential $\phi=0$, and the magnetic field is uniform along the internal directions which have the geometry of  a perfect sphere (\ref{sphericalCompactification}).  Since the gauge symmetry is unbroken everywhere, there are no vortices involved in these solutions.  These compactifications can exist for any value of $\beta$, on a curve of the $\gamma-\beta$ plane  given by the  constraint $|m|\gamma \beta^{1/2} = 2$, where $m$ is the number of magnetic flux quanta.  At the critical coupling, $\beta=1$, the solutions admit an embedding in $\cN=1$  supergravity and are BPS in the sense that they preserve $1/2$ of the supersymmetries.
\item \emph{Critical-critical  string configurations, 
$(n,n)$}:  This type of compactifications, which have a cigar-type of geometry,  only exist as an exact solution to the static equations of motion at the critical coupling, $\beta=1$, with $\gamma=1/|n|$.  As shown by Linet \cite{Linet:1990fk} at this point the inter-string distance becomes a flat direction, however we have found that this distance  has a minimum value  where the configuration reduces to the Salam-Sezgin spherical compactification.  Moreover, these solutions admit an embedding in $\cN=1$ supergravity theory where they preserve $1/2$ of the supersymmetries. As we argued in section \ref{numericalProperties}, away from the critical coupling these configurations cannot exist as static solutions, and would either collapse towards the spherical compactification ($\beta<1$), or decay into two critical strings separated by an infinite distance ($\beta>1$).   Nevertheless, we have argued that these configurations  can still  be regarded as approximate solutions to the static equations of motion 
 for any value of $\beta>1$ provided the inter-string distance is large compared to the size of the  core.
\item \emph{Supercritical-subcritical string configurations, $(n_1,n_2)$ with $n_1> n_2$}: The internal space of these compactifications is warped and exhibits an egg-shaped geometry. We have only found this type of solutions  in the regime $\beta>1$, and thus they can never be BPS in the sense of supersymmetry.  

Each particular solution $(n_1,n_2)$  exists  on a different curve of the $\gamma-\beta$ plane, $\gamma_{n_1,n_2}(\beta)$,  which is bounded above by {\small $2/(n_1+n_2)$}, and  below by the curve associated to a spherical compactification with the same total flux $n_1+n_2$. These solutions can reduce to the  Salam-Sezgin spherical compactification in two different limits: on the one hand, close to the critical coupling $\beta\to 1$, and on the other hand,  configurations  of the form $(m-1,1)$ also approach the spherical compactification  when the magnetic flux is large $m\to \infty$. 
 \end{itemize}

It would be interesting to uplift the present model to a $6d$ theory. We expect most of these solutions to
survive in a higher dimensional setting, although the details of the theory and its matter contents may be
different. Finally, we should also explore the connection of the solutions presented here to more general
time dependent solutions where the strings (or branes in a higher dimensional case) appear to be inflating.
We hope to report on these issues in a future publication \cite{next}.\\

\begin{acknowledgments}

We are grateful to B. Hartmann, J. Senovilla, R. Vera and J.M. Aguirregabiria for very useful discussions.  J.J.B.-P. is supported by IKERBASQUE, the Basque Foundation for Science. B.R. acknowledges financial support from the Basque Government grant (BFI-2011-250). KS gratefully acknowledge support within the framework of the Deutsche Forschungsgemeinschaft (DFG) Research Training Group 1620 {\it Models of gravity}, and the financial support from the University of the Basque Country UPV/EHU via the programme ``Ayudas de Especializacion al Personal Investigador (2012)''.  JU acknowledges financial support from the University of the Basque Country UPV/EHU (EHUA 12/11).  We acknolwedge support from the Basque Government  (IT-559-10), the Spanish Ministry (FPA2009-10612), and the Spanish Consolider-Ingenio 2010 Programme CPAN (CSD2007-00042).
\end{acknowledgments}

\appendix

\section{Conventions}
\label{conventions}
We work in natural units  $\hbar=c=1$ so that the reduced Planck mass reads $M_p^{-2} = 8 \pi G$, and the space-time metric is chosen  to have signature $(-,+,+,+)$. We define the Ricci tensor  following the  conventions from \cite{Dvali:2003zh}. With the ansatz (\ref{lineelement}) the only non-vanishing components of $R_\mu^\nu$ are
\begin{eqnarray}
R_0^0=\frac{(LNN')'}{N^2 L} \ \ , \ \ R_{r}^{r} = \frac{2N''}{N}+\frac{L''}{L} \ \ \ , \ 
\ R_{\varphi}^{\varphi}= \frac{(N^2 L')'}{N^2 L} \ \ \ , \ \ \ R_z^z=R_0^0,
\end{eqnarray}
where the prime denotes the derivative with respect to $r$. The energy momentum tensor is defined as 
\be
T_{\mu\nu} = -2 \frac{\pd \cL_m}{\pd g^{\mu \nu}} + g_{\mu \nu} \cL_m,
\ee
and,  denoting its trace by  $T = T_\mu^\mu$, the Einstein equations take the following form
\be
R_{\mu \nu} = - M_p^2 \left( T_{\mu \nu} -\ft12 g_{\mu \nu }T \right).
\ee
On section \ref{sugraSection} the chirality of the fermions is indicated  by the subscripts $R$ and $L$:
\be
  \chi_{R} = \ft12 (1 - \gamma^5) \chi_{R} \quad \qquad \chi_{L} = \ft12 (1 + \gamma^5) \chi_{L}.
\ee

\section{Space-time matching}
\label{matching}

In this Appendix we will study the matching between two space-times which are static, cylindrically symmetric, and invariant under boosts along the axis of symmetry. Our results show that, in addition to the matching conditions presented in the main body of the paper (\ref{metricMatching1}), it is possible to construct  a configuration of two parallel critical strings where inter-string distance varies at a constant rate $\pd _t r^* =const$.  \\

 We will use the same notation as in section  \ref{matchingSection}. The most general ansatz for the line elements in both patches, which is consistent with these symmetries, is written in equation (\ref{elementNS}) using polar coordinates. Following \cite{Mars1993},  the matching is done identifying the boundaries of the  North and South patches, which are defined by two generic surfaces, $\Sigma^N$ and $\Sigma^S$ respectively, of the form
\bea
\Sigma^{N}:\quad&& \xi^N =\{ \; t(\tau, \zeta, \phi),\;  z(\tau, \zeta, \phi), \;  r(\tau, \zeta, \phi),\;\varphi(\tau, \zeta, \phi)\; \}, 
 \\
\Sigma^S:\quad&& \xi^S =\{\, \tilde t(\tau, \zeta, \phi),\;  \tilde z(\tau, \zeta, \phi), \;  \tilde r(\tau, \zeta, \phi),\; \tilde \varphi(\tau, \zeta, \phi) \;\}. 
\eea

Formally, $\xi^N$ and $\xi^S$ are a couple of diffeomorphisms which map points from an abstract surface $\sigma$ of dimension $3$  to the boundaries   $\Sigma^N$ and $\Sigma^S$ where we use a local coordinate system denoted by $\{y^a\} = \{\tau,\nmu , \phi\}$, which take the values  $\tau, \zeta \in \mathbb{R}$ and $\phi \in [0, 2 \pi)$.  In this approach the matching  surfaces are unknown a priori, and thus they have to be determined together with the matching conditions that have to be satisfied by the metric across the boundaries. Note that these diffeomorphisms also  induce maps from the tangent space of $\sigma$ to the tangent space of the boundaries    
\be
\vec e_a^{\, N} \equiv \xi_*^N(\pd_a) = \frac{\pd x^\mu}{\pd y^a} |_{\Sigma^N} \; \pd_\mu \quad  \qquad \text{and} \qquad \quad  \vec e_a^{\, S} \equiv \xi_*^S(\pd_a) = \frac{\pd \tilde x^\mu}{\pd y^a} |_{\Sigma^S} \; \pd_{\tilde \mu},
\label{pullBacks}
\ee
where the sets $\{\vec e_a^{\, N} \}$ and  $\{\vec e_a^{\, S}\}$ form a basis of the tangent space of their respective surfaces. 

\addtocontents{toc}{\protect\setcounter{tocdepth}{1}}
\subsection{Symmetry preserving matchings}
\addtocontents{toc}{\protect\setcounter{tocdepth}{2}}

In order for the final  space-time geometry to respect the symmetries of the individual patches, the matching itself has to be consistent with them.  This will lead, as we shall see, to   constraints in the form of the embedding functions $\xi^N$ and $\xi^S$.\\

  The relevant symmetries of the North patch are generated by the following killing vectors (with analogous expressions for the South patch)
\be
\vec  k_z = \pd_z,\qquad  \vec   k_\varphi = \pd_\varphi,\qquad \text{and} \qquad \vec  k_b = t \, \pd_z - z \, \pd_t,
\ee  
which are associated respectively to  translations along the $z$ direction, azimutal rotations and boosts along the symmetry axis $z$. Note that we will not  impose invariance under time translations, so that we can also  study   time dependent  matchings.  For the matching to be consistent with these symmetries we have to request the surfaces $\Sigma^N$ and $\Sigma^S$ to be everywhere tangent to the corresponding  killing vector fields  \cite{Vera2002,rverathesis}. In order to fulfill this condition  a subset of the basis vectors $\vec e_a^{\, N}$ (\ref{pullBacks})  must  be identified with  a set of killing vectors  which generate the corresponding symmetry algebra \cite{Vera2002}.  On the one hand, it is possible to argue  that, due to the intrinsic properties of the killing vector\footnote{It generates a compact symmetry of period $2 \pi$ which leaves the $z$ axis invariant.} $\vec k_\varphi$, it can be identified with $\vec e_1^{\, N}$, the image of the axial vector  $\pd_\phi$  th
 rough $\xi^N_*$. On the other hand,  the image of $\pd_\zeta$ can be a general space-like combination of all the killing vector fields which commutes with $\pd_\varphi$
\be
 \vec e_2^{\, N}\equiv  \xi_*^N(\pd_\phi )=  \vec k_\varphi,  \qquad \qquad
\vec e_3^{\, N}\equiv  \xi_*^N(\pd_\zeta)=  a \, \vec k_t + b \, \vec  k_z + c  \, \vec k_\varphi + d\, \vec k_b,
\label{symmetryConstraints1}
\ee
where $a$, $b$, $c$ and $d$ a priori are arbitrary constants.  The condition that $\vec e_3^{\, N}$ is space-like everywhere, implies that the boosts can not appear in the linear combination combination i.e. $d=0$. Moreover, choosing conveniently  the coordinates on $\sigma$,  and using the freedom to reparametrize $t$ and $z$ in the North patch which is left after imposing the ansatz (\ref{elementNS}), it is always possible to set $b=1$ and $a = c = 0$.  Since the metric is diagonal (\ref{elementNS})  the basis vector $\vec e_1^{\, N}$ can be chosen, without loss of generality, to be orthogonal to the other two. Taking this into account and comparing equations (\ref{pullBacks}) with (\ref{symmetryConstraints1}), we find a set of differential equations for the  embedding functions which, in the case of $\xi^N$, have the general solution
\be
\xi^N :\qquad  t =  f(\tau), \qquad z= \nmu, \qquad r=r_N(\tau),\qquad \varphi = \phi,
\label{EmbeddingNorth}
\ee
with $f(\tau)$ and $r_N(\tau)$ being arbitrary functions of the parameter $\tau$. Thus, the basis vector $\vec e_1^{\, N}$ takes the form
\be
\vec e_1^{\, N} \equiv  \xi_*^N(\pd_\tau )  =[ \dot{f} \, \partial_t + \dot{r}_N\, \partial_r]_{\Sigma^N},
\label{symmetryConstraints2}
\ee
where  the dot denotes a derivative respect to $\tau$. For the South patch it is possible to find  analogous expressions for the embedding functions $\xi^S$ \be
\xi^S :\qquad   \tilde t = \, h(\tau), \qquad \tilde z=  \nmu, \qquad \tilde r=\tilde r_S(\tau),\qquad \tilde \varphi = \phi + \tilde c \, \zeta,
\label{EmbeddingSouth}
\ee
and for the basis vectors $\{\vec e_a^{\, S}\}$  in terms of the arbitrary functions $h(\tau)$ and $r_S(\tau)$
\be
\vec e_1^{\, S} = [ \dot{h} \,\partial_{\tilde t} + \dot{r}_S\,\partial_{\tilde r}]_{\Sigma^S} \qquad
\quad  \vec e_2^{\, S}=    \vec k_{\tilde \varphi},  \qquad\quad 
\vec e_3^{\, S}=   \vec  k_{\tilde z} + \tilde c\,  \vec k_{\tilde \varphi}.
\ee

 In order to simplify the expression for $\vec e_3^{\, S}$ we have used the freedom to reparametrize $\tilde t$ and $\tilde z$ in the South. However, since we have already used the freedom to parametrize $\sigma$ to simplify (\ref{symmetryConstraints1}), in the South patch it is not possible  to set the constants $\tilde c$ to zero.\\

  In order to compute the matching conditions in the next subsection we will also need the normal vectors to the surfaces\footnote{We restrict ourselves to timelike matching hypersurfaces. If we wanted them to became null somewhere, instead of the normal vector we should use some vector transverse to the matching hypersurface (a {\it rigging} vector)\cite{Mars1993}.} $\Sigma^{N,S}$, which are given by
\be
\vec{n}^N=-[\frac{\dot{r}_N}{N} \, \partial_t+ \dot{f}\, N \, \partial_r]_{\Sigma^N}, \quad \qquad 
\vec{n}^S =[  \frac{\dot{r}_S}{\Nm}  \partial_{\tm}+ \dot{h}\, \Nm  \, \partial_{\rrm}]_{\Sigma^S}.
\ee
With all this results at hand   the matching can  be defined now  by the identification of the boundary surfaces $\Sigma\equiv \Sigma^N\sim \Sigma^S$,  through $\xi ^N$ (\ref{EmbeddingNorth})  and $\xi^S$ (\ref{EmbeddingSouth}), of  their  tangent space   $\vec e_a^{\, N} \sim \vec e_a^{\, S}$, and of the normals $\vec n^N \sim  \vec{n}^S$.

\addtocontents{toc}{\protect\setcounter{tocdepth}{1}}
\subsection{Matching conditions}
\addtocontents{toc}{\protect\setcounter{tocdepth}{2}}   

The preliminary matching conditions ensure the continuity of the metric across the matching surface $\Sigma$, and are necessary in order to have a globally well defined metric.   These conditions  are equivalent to imposing the equality of the inner products of the tangent basis with respect to the metric in the corresponding patch. In this case, these conditions yield
\be
L^2 =  \Lm^2, \qquad \quad 
N^2 =  \Nm^2, \qquad \quad 
(\dot{f}\, N)^2 -\dot{r}^2_N =  (\dot{h}\, \Nm)^2 -\dot{r}^2_S,\label{htt}\\
\ee
together with $\tilde{c}=0$.  The remaining matching conditions are obtained imposing the equality of the second fundamental forms of both patches $(H_{ab} = -n_\nu e_a^\nmu \nabla e_b^\nu)$ at $\Sigma$, and  ensure that there is no energy concentration on the boundary surface.   Assuming $N,L,\Nm,\Lm \geq 0$, they read
\begin{eqnarray}
H_{\tau \tau}&:&N^2N' \dot{f}^3 - 2N'\dot{f}\dot{r}^2_N+N(-\dot{r}_N\, \ddot{f}+\dot{f}\, \ddot{r}_N) = \nonumber\\
&&-\left[ \Nm^2 \Nm ' \dot{h}^3 - 2\Nm '\dot{h}\dot{r}_S^2+ \Nm(-\dot{r}_S\, \ddot{h}+\dot{h}\, \ddot{r}_S)\right],\nonumber\\
H_{\phi \phi}&:&N\dot{f}L' = -  \Nm \dot{h}\Lm',\nonumber\\
H_{\nmu \nmu} &:& N'\dot{f} = - \Nm' \dot{h} \label{Htt}.
\end{eqnarray}

Then,  the most general embedding functions $\xi^N$ and $\xi^S$ consistent with system of equations (\ref{htt}) and (\ref{Htt}) can be written in the following way (with $\dot f/ \dot h>0$):
\begin{eqnarray}
 N\,  \dot f =A(\tau) \cosh( \alpha (\tau)) \;, && \; \dot r_N=A(\tau) \sinh( \alpha(\tau)),\nonumber \\
 \tilde N \, \dot h=A(\tau) \cosh(\alpha_0-\alpha(\tau))\;, && \; \dot r_S=A(\tau) \sinh(\alpha_0- \alpha(\tau)),
\end{eqnarray}
where $A(\tau)$ and $\alpha(\tau)$ are arbitrary functions of $\tau$, $\alpha_0$ is a real constant. 
Thus, the matching sufaces $\Sigma^{N,S}$ are two cylinders centered in $r=0$ and $\tilde r=0$, with  time dependent radii:
\be
\pd_t \, r_N \equiv \dot r_N / \dot f=  \tanh(\alpha(\tau))/N, \qquad \quad \pd_{\tilde t}\,  r_S\equiv \dot r_S /\dot h=\tanh(\alpha_0 -\alpha(\tau))/\tilde N.
\ee

We can  distinguish two different cases:
\begin{itemize}
\item \emph{ Supercritical-subcritical matching}:  in this case  at least one of the metric profile functions $L'$, $\tilde L'$, $N'$ or $\tilde N'$ is non-zero at the matching surface. In that case it can be shown that the constant $\alpha_0$ must be vanishing\footnote{Taking derivatives of  (\ref{htt}) with respect to $\tau$, (with  $\dot N = N' \, \dot r_N$ and $\dot L = L' \, \dot r_N$) and comparing the result with (\ref{Htt}), leads to a constraint on $\dot f$, $\dot h$ and $\dot r_{N,S}$. }. 
Thus the surfaces $\Sigma^{N,S}$ move together  over the compact space  with arbitrary speeds $\pd_t \, r_N  =-\pd_t\,  r_S$, leaving  the inter-string distance $r^*$  unchanged. Since the matching surface is not physical, all these matchings  are equivalent to each other, and  in particular setting $\alpha(t)=0$ we recover the  matching (\ref{metricMatching1}), where the radii $r_N$ and $r_S$ are time independent parameters.

\item \emph{Critical-critical matching}: here all the quantities $L'$, $\tilde L'$, $N'$ and $\tilde N'$ are  zero at the matching surface and then the equations  admit and  new solution in addition to the previous one. Requiring  $t$ and $\tilde t$ to run at the same rate ($ \dot f = \dot h$) then the solution is obtained setting $\alpha(\tau) = \alpha_0/2$. In this case the matching surfaces $\Sigma^{N,S}$ move at a  constant speed $\pd_t \, r_N  =\pd_t\,  r_S$,  from  the center of the strings. Depending on the sign of the velocities the strings move towards each other ($r^*$ decreases) or away from each other ($r^*$ increases) at a constant rate.  Here again the matching conditions for the metric are given by (\ref{metricMatching1}). 
\end{itemize}

\section{Perturbative analysis near the BPS limit}
\label{analyticSolutions}

 In this Appendix we  use perturbation theory techniques to identify solutions of type $(n_1,n_2)$ for values of $\beta$ near the critical coupling: $\beta=1+s\, \epsilon$ . For convenience we take the perturbation parameter to be positive, $1\gg\epsilon>0$, with the constant  $s$ taking the values $\{-1,0,1\}$ in order to account for the cases  $\beta<1$, $\beta=1$ and $\beta>1$ respectively.  
In section \ref{LinetCompactification}  we proved that at the critical coupling the only possible solutions are the spherical compactification (\ref{sphericalCompactification}) and the $(n,n)$ compactification found by Linet. Therefore  we take as the zero order solution the spherical compactification at critical coupling with the same total flux as the $(n_1,n_2)$ solution, i.e. $m=n_1+n_2$, and the gravitational coupling satisfying  $\gamma=2/|m|$.
 As we discussed  in section \ref{numericalResults}, regular compactifications induced by  supermassive cosmic string only exist on a curve of the $\gamma-\beta$ plane, $\gamma_{n_1,n_2}(\beta)$ and thus     
we also perturb the value of  $\gamma$ 
\be
\gamma_{n_1,n_2}(1 + s \epsilon) = \ft{2}{|m|} +\ft{\gamma_1}{|m|}  \epsilon +  O(\epsilon^3) .
\ee

For the profile functions we will  introduce into the equations of motion an ansatz of the form
\be
f(r) =\epsilon^{1/2} \sum_{k=0}^\infty f_{\ft12 +k}(r) \epsilon^k,\quad v(r) =\sum_{k=0}^\infty v_{k}(r) \epsilon^k,\quad L(r) = \sum_{k=0}^\infty L_{k}(r) \epsilon^k, \quad N(r) = \sum_{k=0}^\infty N_{k}(r) \epsilon^k,
\label{perturbation}
\ee
and then solve the resulting differential equations for each order in $\epsilon$.  Here $v_0$,  $L_0$ and $N_0$ are the profile functions of the zero order solution, (\ref{sphericalCompactification}). At this point it is convenient to change the matching scheme  in order to simplify the calculations. Instead of the definition  (\ref{boundary1}) for the boundary,  we will use $L(r_N)=0$, which implies that $r_N= r^*$ and $r_S =0$.
With this choice the matching conditions  (\ref{metricMatching1}-\ref{gaugeMatching2}) together with the boundary  conditions at $\tilde r=0$ (\ref{boundaryConditions}) translate into a new set of boundary conditions at $r=r^*$:
\be
f(r^*)=0, \qquad v(r^*)=n_1+n_2, \qquad L'(r^*)=-1, \qquad N'(r^*) = 0,
\label{newBoundaryConditions}
\ee
which ensure the regularity of space-time and of the scalar field configuration at $r^*$. Note that the point where the profile function $L(r)$ vanishes, also changes when we switch on  the perturbation $r^* \approx \pi r_0 + \epsilon r_1 + \ldots$, with  $r_0\equiv\sqrt{|m|/2}$, and therefore the boundary conditions    (\ref{newBoundaryConditions})  have to  be changed accordingly. For example,  to leading order in $\epsilon$ we find the following boundary conditions for the perturbed  profile functions at $r= \pi r_0$:  
\be
L_1(\pi r_0) = r_1, \qquad L'_1(\pi r_0) = 0, \qquad f_{1/2} (\pi r_0 ) = 0, \qquad  v_1(\pi r_0) = 0 , \qquad N_1(\pi r_0) = 0.
\label{boundaryVF1}
\ee

In the following sections in this Appendix, we will use this analysis to search for the perturbative expansion of new solutions in different sectors of the parameter space.

\addtocontents{toc}{\protect\setcounter{tocdepth}{1}}
\subsection{Zero-mode in $(n,n)$ configurations in the BPS limit}
\addtocontents{toc}{\protect\setcounter{tocdepth}{2}}

Plugging the previous ansatz for the parameters and   (\ref{perturbation}) into the system of equations (\ref{systemE2}) we obtain a  set of second order differential equations for the perturbations. However, it is not difficult to check that the equation for  $f_{1/2}(r)$ can be integrated once, leading the first order differential equation 
\be
f'_{1/2} - L_0^{-1} \, (n - v_0) f_{1/2}=0,
\label{perturbed1/2}
\ee
which can be solved analytically giving, in the case $n_1=n_2\equiv n $ 
\be
f_{1/2}(r) =  f_{(n,n)} \, \sin^{|n|} ( r/r_0).
\ee

After solving the resulting equations at order one,  the boundary conditions (\ref{boundaryVF1})  determine the correction to the coupling, $\gamma_1=-1$. It can also be shown that the profile functions $v_1$, $L_1$ and  $N_1$ scale as $f_{(n,n)}^2$, which remains  free to order  one. At the next order, the equation and the boundary conditions for $f_{3/2}$, fix the value of $s=0$,  implying that those solutions can only be found at the critical coupling, $\beta=1$.   This result   allows  to find an analytic expression for the first order corrections to the profile functions: 
\be
v_1(r) = - |n| \,a_n(r)  \sin(\ft{r}{r_0}), \ \ L_1(r) =  - r_0   \, a_n(r) \cos(\ft{r}{r_0}),  \ \ a_n(r) \equiv f_{(n,n)}^2\int_0^{r/r_0} du \, (\sin u)^{2n},
\label{perturbedBPSsols}
\ee
with  the gravitational potential  left unperturbed  $N_1(r)=0$. Since in this case the $\epsilon$ is not linked to the parameters of the theory, it becomes an arbitrary constant of the solutions, in other words it is  associated to a zero-mode, and moreover  the constant $f_{(n_1,n_2)}$ can be absorbed in the perturbation parameter. Thus we find the following relation between the maximum value of $f(r)$ with  the inter-string distance $r^*$: 
\be
r^* = \pi  r_0 ( 1+ \ft {(2 n)!}{(2^n n!)^2}   {\epsilon})  + O(2) \qquad \Longrightarrow \qquad   f_{max} (r^*)  = \ft {2^n n! }{\sqrt{(2 n)!} }  \, (\ft{r^*}{\pi  r_0  } - 1)^{1/2}. 
\ee

The profile gauge and metric functions  obtained using this method agree perfectly with the results obtained with the full  numerical simulations up to inter-string distances of order $r^*\sim3.3$. The scalar profile function obtained  in this way is not as accurate as the other profiles due to the fact that at leading order the   the correction for  $f(r)$ does not take into account the variation of the boundary point $r^*$ (\ref{boundaryVF1}).

\addtocontents{toc}{\protect\setcounter{tocdepth}{1}}
\subsection{Approximate analytic $(n_1,n_2)$ configurations near the BPS limit}
\addtocontents{toc}{\protect\setcounter{tocdepth}{2}}     

Proceeding as in the last subsection we now search for configurations of the form $(n_1,n_2)$ near the BPS limit. In this case the  equation for the leading order correction to $f(r)$ is still  (\ref{perturbed1/2}) which can be solved by
\be
f_{1/2}(r) = f_{(n_1,n_2)}\, \left(\ft{m}{2 n_1}(1- \cos(r /r_0))\right)^\ft{n_1}{2} \, \left(\ft{m}{ 2n_2}(1+ \cos(r /r_0))\right )^\ft{n_2}{2}.
\label{perturbedBPSsols1}
\ee

As in the previous case the equations at order one for the  profile functions $v_1$, $L_1$ and  $N_1$, together with the boundary conditions   (\ref{boundaryVF1}) fix the value of $\gamma_1 =-1$ leaving  the parameter $f_{(n_1,n_2)}$ still arbitrary.  This freedom is lifted at the next order where the equation and the boundary conditions for $f_{3/2}$, fix simultaneously the value of $f_{(n_1,n_2)}$ and  $s=1$, implying that these solutions only exist for $\beta>1$. Then, from (\ref{perturbation}) and (\ref{boundaryVF1}) we see that the  maximum of the scalar profile function and the inter-string distance  depend on $\beta$ in the following way   
\be
f_{max}(\beta) = f_{(n_1,n_2)}  \, (\beta-1)^{1/2}, \qquad r^* (\beta) = \pi r_0 + L_1(\pi r_0)\, (\beta -1).
\label{condensateSize}
\ee

The profile functions $L(r)$, $v(r)$ and $N(r)$  obtained to leading order in perturbation theory, fit very well the results of the full numerical simulations up to  values of $\beta \sim 1.01$. As in the case of the critical-critical configurations the scalar profile function calculated to leading order shows slight differences with the numerical results. Again  this is due to the fact that the leading order correction of $f(r)$ is $\cO(\ft12)$ while the leading order perturbation for the other fields is $\cO(1)$.

\section{Critical-critical configurations in the large$-n$ limit.}
\label{largeNAppendix}

In this section we  derive an analytic expression for an $(n,n)$ configuration in the limit of large magnetic flux, that is  $n\to \infty$. We will proceed following  \cite{Achucarro:1993bu}, where the authors study the large winding number limit for flat space Abelian-Higgs vortices. More technical details about the method we use can be found in \cite{holmes}.\\

Inside the core of a gravitating cosmic string with a large winding number  $n$, the system of equations of motion (\ref{systemNO1}-\ref{systemE2}) at leading order in $1/n$ reduces to  
\bea
\gamma n\,  v'  N^2 = L, &\qquad& \xi' - L^{-1} (1- v) \, \xi \ = \ 0, \\
(LNN')' =\frac {1} {2 \gamma n^2}  (1 - \gamma^2 n^2 \beta N^4) \frac{L}{N^2}=0, &\qquad& (N^2 L')' =\frac {-1} {2 \gamma n^2}  (1 + \gamma^2 n^2 \beta N^4) \frac{L}{N^2}
\label{E1}
\eea
\begin{figure}[t]
\centering
\subfigure[ Gravitational potential $N(r)$]{
   \includegraphics[width=0.4\textwidth]{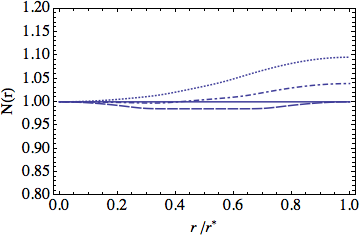}
}
\hspace{2.2cm}
\subfigure[Metric profile function $L(r)$]{
   \includegraphics[width=0.4\textwidth]{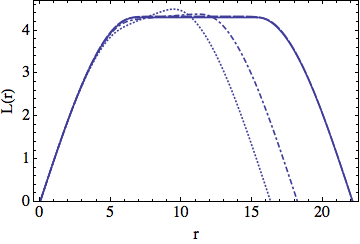}
}
\subfigure[Condensate size $f(r)$]{
   \includegraphics[width=0.4\textwidth]{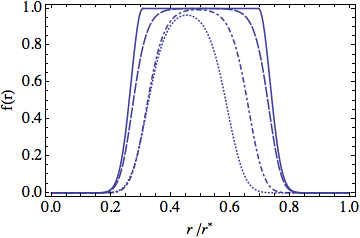}
}
\hspace{2.2cm}
\subfigure[Magnetic field density $B(r)$]{
  \includegraphics[width=0.4\textwidth]{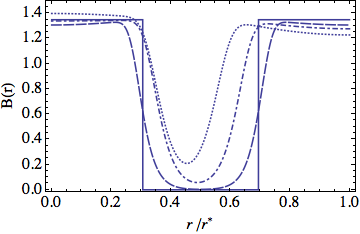} 
}
\caption{\footnotesize{Profile functions of different configurations with a total magnetic flux $2n=50$  at  $\beta = 1.818$,  together with the corresponding large$-n$ approximation (solid line). We show the splittings  
$(25,25)$, $(26,24)$ and $(29,21)$ with  dash-dotted,  dashed,  and dotted lines respectively.}}
 \label{largeNprofiles}
\end{figure}

In order to derive these equations we have performed the change of variables $f = \xi^n$, and then used the fact that inside the core  $\xi <1$. In particular,  this implies that to leading order  all the terms where $f(r)$ appears are vanishing. Note that the equation for the gauge profile function has been integrated once, thus  it is also necessary to fix an integration constant using the boundary condition for the gravitational potential   $N(0)=1$, taking into account  that critical strings satisfy $B(0) =1/\gamma n$  (\ref{BNrelation}). The $rr-$component of Einstein's equations provides an extra constrain on the fields \cite{Christensen:1999wb}, which in this case reduce to 
\be
 N' \left(2 L' N+ N' L\right) =\frac {1} {2 \gamma n^2} \left(1 -  \gamma^2 n^2 \beta N^4 \right)\frac{L}{N^2}.
\label{constraint}
\ee 

Since we are interested in a critical-critical configuration  the  profile  functions must satisfy the boundary conditions  at the outer limit of the core, $r_{1/2}$,
\be
f'(r_{1/2}) = 0, \qquad v (r_{1/2}) = n, \qquad  L'(r_{1/2}) = 0, \qquad N'(r_{1/2}) = 0.
\label{contraintE}
\ee

 It is possible to show that the equations of motion (\ref{E1}), the constraint (\ref{contraintE}) and the previous boundary conditions can only be solved provided the parameters satisfy $n \gamma \sqrt{\beta} = 1$, which coincides  precisely with the constraint equation of the  spherical compactifications (see section \ref{sphericalCompactificationSection}).  Actually  we   recover the same solution as for the  spherical compactification (\ref{sphericalCompactification})  inside the core of the strings, i.e.  in the interval   $r\le r_{1/2} = \frac{\pi}{2} r_0$, with the total flux given $m=2 n$. Finally the equation for the scalar field can be solved  giving
\be
\quad f(r) = \sin^n(r/r_0). 
\ee

In Fig. \ref{largeNprofiles} we compare the large$-n$ approximation of the profile functions with the results of a numerical simulation with total flux $2 n=50$ at $\beta = 1.818$. It can be seen that inside the core the analytic expressions fit well the qualitative features and scales of the numerical $(n,n)$ solution, and that the differences become more evident in the transition region,  $r \sim r_{1/2}$. In particular note that the metric profile function obtained with the two different methods are indistinguishable, implying that to a very good approximation the geometry of these configurations near the string center is essentially a spherical cap. Moreover, the numerical analysis shows that for this solution $ n\,  \beta^{1/2}\, \gamma_{25,25} = 1.04$, in good agreement with the prediction that in the large$-n$ limit the parameters should satisfy the same constraint as in the case of the spherical compactification.

\bibliography{supermassive}

\end{document}